# Vulnerabilities and Attacks Against Industrial Control Systems and Critical Infrastructures

Georgios Michail Makrakis *, Constantinos Kolias *, Georgios Kambourakis †, Craig Rieger ‡, and Jacob Benjamin §

*Department of Computer Science, University of Idaho, Idaho Falls, ID, 83402 USA, (e-mail: gmakrakis@uidaho.edu, kolias@uidaho.edu)

†European Commission, Joint Research Centre (JRC), 21027 Ispra, Italy, (e-mail: georgios.kampourakis@ec.europa.eu)

‡Idaho National Laboratory (INL), Idaho Falls, ID, 83402, USA, (e-mail: craig.rieger@inl.gov)

§Dragos, Hanover, MD 21076 USA, e-mail: (jbenjamin@dragos.com)

## Abstract

Critical infrastructures (CI) and industrial organizations aggressively move towards integrating elements of modern Information Technology (IT) into their monolithic Operational Technology (OT) architectures. Yet, as OT systems progressively become more and more interconnected, they silently have turned into alluring targets for diverse groups of adversaries. Meanwhile, the inherent complexity of these systems, along with their advanced-in-age nature, prevents defenders from fully applying contemporary security controls in a timely manner. Forsooth, the combination of these hindering factors has led to some of the most severe cybersecurity incidents of the past years. This work contributes a full-fledged and up-to-date survey of the most prominent threats against Industrial Control Systems (ICS) along with the communication protocols and devices adopted in these environments. Our study highlights that threats against CI follow an upward spiral due to the mushrooming of commodity tools and techniques that can facilitate either the early or late stages of attacks. Furthermore, our survey exposes that existing vulnerabilities in the design and implementation of several of the OT-specific network protocols may easily grant adversaries the ability to decisively impact physical processes. We provide a categorization of such threats and the corresponding vulnerabilities based on various criteria. As far as we are aware, this is the first time an exhaustive and detailed survey of this kind is attempted.

## I. Introduction

CRITICAL infrastructures (CI) are comprised of systems and assets so indispensable for the proper function of society that their deterioration will surely prove detrimental to public health, national security, and economic well-being. Such systems cover multiple facets of our everyday lives, but water, energy, communications, and transportation, are considered among the most vital sectors. Security of CI has always been in the epicenter of thorough assessments, but until today it was mainly geared to prevent random accidents and man-made physical assaults. Today, due to the increasingly more significant role of IT systems in the operation of CI, such environments have also become the subject of cyber threats.

An Industrial Control System (ICS) can conceptually be subdivided into the IT and Operational Technology (OT) domains. The IT portion is centered on providing all services that support the business operations. It is comprised of workstations, servers, and databases, all of which are interconnected using IP-based networks. The OT portion focuses on the operational aspects of machinery. The main components of OT systems are domain-specific devices such as Programmable Logic Controllers (PLCs) and Variable-Frequency Drives (VFDs).

In their majority, ICS consist of components for which their hardware, software, and networking elements, are optimized to have prolonged life of many decades. Interestingly, despite their critical nature, many ICS devices do not inherently support any cybersecurity mechanisms.

It can be argued that today, the vast majority of cybersecurity practitioners have only a superficial knowledge of ICS. Yet, due to its pivotal role in CI, it is worth putting ICS security under the magnifying lens. Unlike modern IT systems, ICS operate beyond the boundaries of cyberspace and are rather entangled with the physical domain. For this reason, an anomaly caused by a security breach may not only inflict significant economic losses or loss of privacy, which are the typical worst-case scenarios in pure IT systems. In the most dreadful scenarios, such violations may result in wide-reaching environmental destruction or put the safety of citizens at risk.

Naturally, the potential of large-scale and high-profile impact makes ICS inside the CI, alluring targets for various adversaries. These actors bear different characteristics than the stereotypical "IT hacker", as they usually have many more resources at their disposal and are driven by motives that range from the mere pursuit of profit, but may expand to applying geopolitical pressure.

At the same time, the security of the CI is a daunting task for multiple reasons. Besides the existence of a large number of legacy equipment and the insecurity of the communications in ICS, which are discussed in the main sections of the paper, the complexity of the systems requires operators with a deep understanding of the multiple domains that constitute CI; typically, a CI is comprised of a myriad of ICS devices of different types. Human engineers are required to have full knowledge of the operational details not only for the devices directly involved in their assigned tasks, but also a well-rounded overview of the system. However, at this point, a dichotomy exists as IT and OT personnel appear to have disjoint training backgrounds. This results in certain aspects of the system, including critical security functions, being viewed by their operators as black-boxes. Therefore, it does not come as a surprise that errors caused by the human factor are still the primary reason behind the majority of the incidents observed in real life.

On top of the aforementioned reasons, one should also take into account the integration of IT and OT realms, a tendency that is observed today lately across virtually all CI sectors. Driven by the desire to evolve the production processes to fit into the broader context of the fourth industrial revolution (Industry 4.0), the paradigm of the Industrial Internet of Things (IIoT) is introduced to assist such evolution with the extensive use of Big Data and Data Mining techniques. Along with this development comes the need for better security, as the systems become more complex and the attack surface broadens. Therefore, the organizations should examine the additional risks introduced by this new class of technology integration, how the requirements in the existing and planned standards will be impacted, and how potential cybersecurity solutions can fit at the very beginning of IIoT implementation.

Through the analysis of real-life incidents, several other factors have been identified and are outlined in subsequent portions of the paper. The purpose of the work at hand is to offer a full-scale survey around the current state of play of ICS and CI security. After examining the related work and defining an adversarial model, we meticulously examine and categorize the vulnerabilities that originate from the potential insecurities of the integrated cyber systems, including the relevant networking protocols and devices. We perform a deep analysis of the most well-known security incidents against such systems, based on the most preeminent information acquired both from academic work and reports/whitepapers created by the overall security industry. We classify the vulnerabilities and attacks based on the adversary's methodology, potential damage, attack impact, and available countermeasures. However, we do not cover any specific work that discusses in depth the aspects of defense tools, SCADA, IIoT, ICS testbeds, and human factor-related threats. For all of the above, we refer the concerned reader to following [1], [2]. [3] [4].

Specifically, the *key contributions* of this work are as follows:

- We detail on the general ICS architecture and its main components in terms of hardware, industry-specific protocols, and common security practices. We also elaborate on the definition of CI and the main threats against it.
- We offer a comprehensive analysis and discussion of the hitherto major ICS and CI security incidents. This enables a comprehensive view of the attackers' tactics, techniques, and procedures. The incidents are further taxonomized based on the type of vulnerabilities that leverage the affected level of the ICS, their outcomes, and the possible mitigation strategies.
- A comprehensive review of the security characteristics of all prominent communication protocols employed in the context of ICS and CI. This line of discussion also elaborates on protocols' vulnerabilities as pinpointed by the relevant literature, and therefore results in common attack types and major challenges towards providing a better security posture.
- An analysis and discussion of the vulnerabilities that exist in ICS-specific devices that have been discovered in academia and how these vulnerabilities are employed against the control process of ICS and CI.

Given the above, vis-à-vis the relevant literature, the current work is the first to our knowledge to not only provide a full-grown, extensive, and contemporary analysis of the major security incidents against ICS and CI, but also to blend this analysis with both the practical and theoretical security shortcomings pertaining to all key operational layers of the ICS. Particular focus is given to the vulnerabilities that affect the layers that are closer to the physical process. This choice is made since IT-related vulnerabilities have been investigated thoroughly in the past, and that ICS-specific issues present unique characteristics worth investigating.

The remainder of this paper is structured as follows. The next section II provides background information about ICS and CI. Section III addresses the related work. The adversarial model is given in section IV. Section V details on major ICS and CI cybersecurity incidents reported over the last few years. The analysis of the incidents also focuses on the reasons why each attack was prosperous. Section VI concentrates on prominent ICS protocols and elaborates on their potential weaknesses as identified by the relevant literature. Section VII discusses the case of vulnerable ICS devices and the repercussions that they can have to the controlled processes. The last section concludes and offers pointers to future work.

## II. Background

This section provides background information regarding the prevailing terms seen in ICS environments. A level of familiarity with all these concepts is necessary to better comprehend the discussions included in the main sections of the paper.

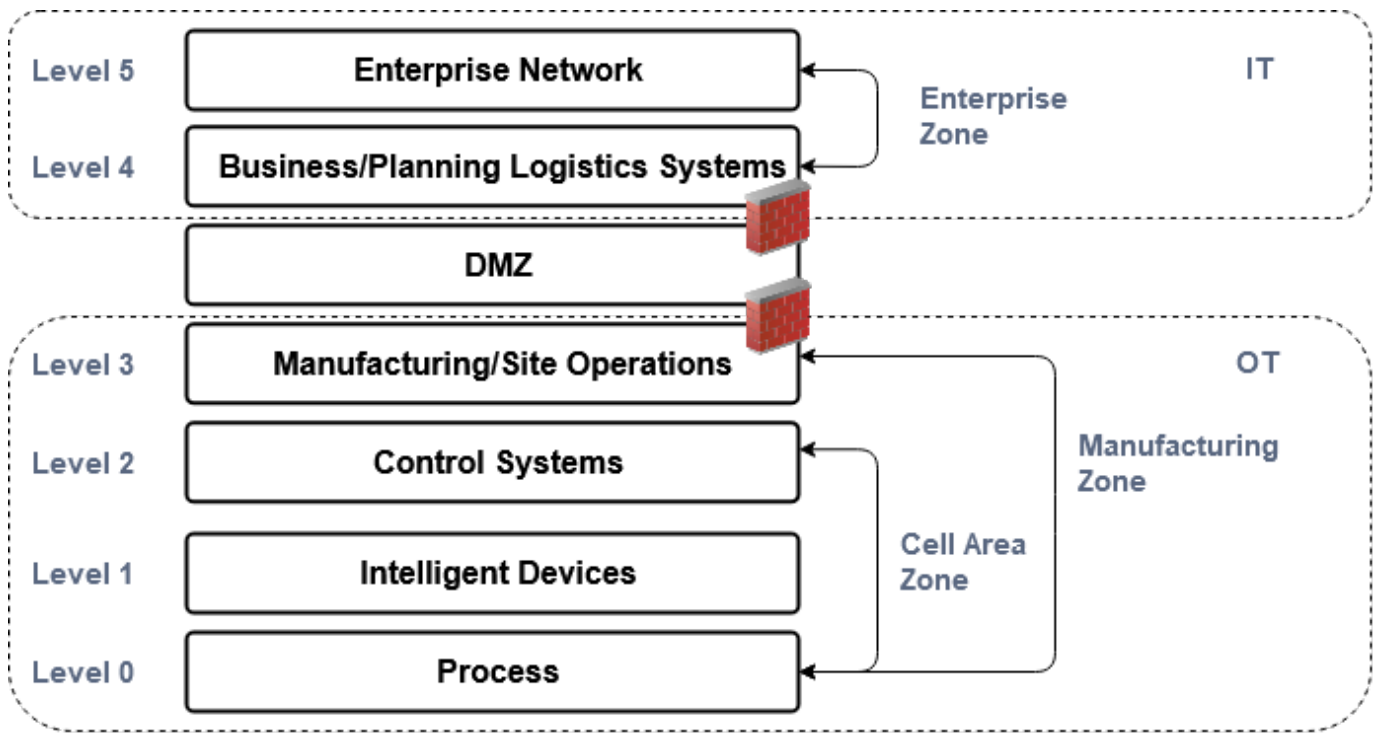

Figure 1: An adaptation of the Purdue Enterprise Reference Architecture by ISA-99

## A. ICS Architecture

When attempting to describe ICS architectures the Purdue Enterprise Reference Architecture (PERA), or simply Purdue Model [5] is usually adopted. The model represents the systems that may be tracked in typical ICS into levels. The model represents the systems that may be tracked in typical ICS into levels can be further grouped into "zones". Each one of these represent a distinct section of functionality offered to the ICS.

A brief explanation of each level of the model follows:

- Level 0 — The actual physical processes that run into the facility. Here, we meet devices, including sensors motors, pumps, and valves, that is, instruments whose main purpose is to provide sensing or actuating capabilities to the system.
- Level 1 — Intelligent devices that sense, monitor, and control the physical processes. Such devices are the Programmable Logic Controllers (PLCs), Proportional-Integral-Derivative (PID) controllers, and the Safety Instrumented System (SIS) controllers.
- Level 2 — Control systems used for supervising and monitoring the physical processes. Among others, this level includes Human-Machine Interface (HMIs) and Engineering Workstations (EWs).
- Level 3 — Manufacturing/Site operations systems used to manage the production workflow for plant-wide control. Devices typically found in this level are the Data Historians, Microsoft Active Directory Domain Controllers and file servers.
- Industrial Demilitarized Zone (DMZ) — Created to prevent the direct communication between IT and OT environments by installing "broker" services. Proxy servers, database replication servers, and remote access servers are typical entities in this extra level.
- Level 4 - Business/Planning logistics systems used to oversee the IT-related activities of the site operations that support the production process. Some of the systems in this level are application servers, email clients and servers and ERP systems.
- Level 5 - The enterprise network used for production and resource data exchange for business-to-business, and business-to-customer purpose services.

A high-level adaptation of the Purdue Model and the main elements of this architecture are illustrated in Figure 1. Based on the above adaptation of the Purdue Model, a typical environment can be subdivided into IT and OT networks. The former comprises conventional PCs, application servers, e-mail servers, ERP systems, etc. Nowadays, we observe a clear convergence of OT and IT network divisions. Therefore, standard IT components can be found in OT realm, such as desktop PCs, and industrial devices communicating via either standard protocols such as TCP/UDP or via industrial protocols as detailed in Section VI. Typical hardware components that comprise the OT network are given in Figure 2.

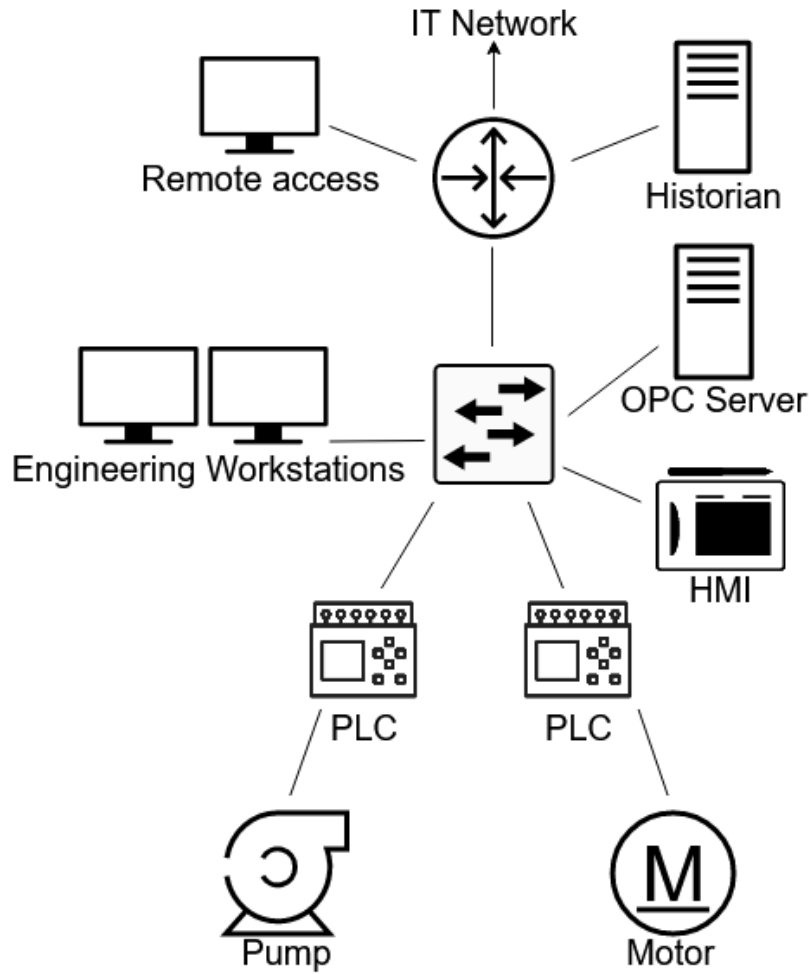

Figure 2: A typical OT network topology

Naturally, by moving downwards in the model, different levels of trust for the underlying devices are established. For example, devices that reside inside the enterprise and business levels have lower trust, the DMZ entities medium trust, and levels 0 to 3 higher trust. All these are based on the restrictions in terms of the installed equipment and software, as well as the physical access to these systems. Naturally, this is also subject to the particular requirements of each sector and facility of interest.

*B. ICS Hardware*

The level 1 of an ICS can include PLCs, Remote Terminal Unit (RTUs), Intelligent Electronic Device (IEDs), and SIS controllers. HMIs belong to level 2, EWs in levels 2 to 3, Data Historians in level 3, remote access servers (or "jump servers") in DMZ, and typical network management devices such as layer 3 and layer 2 switches, routers and firewalls placed in various levels. Other devices can also be present, depending on the requirements and the utilities or products the industrial facility provides. In terms of hardware, the devices in the upper levels of the model (levels 2-3) resemble typical IT devices, e.g., PCs that run MS Windows OS and multicore processor servers with surplus memory. These devices become more common even in advanced-in-age installations. One might also consider that while HMIs were once separate devices, there are now frequently implemented as desktop applications. A detailed description of the aforementioned types of devices remains out of the scope of this paper; however, the interested reader can obtain further information from the work in [6].

Industrial devices in the lower levels of the model, namely levels 1 or 2 have a) much lower hardware specifications, say, a few MHz CPU cycles, a few kilobytes or megabytes of memory, b) run real-time operating systems (RTOS) created for deterministic performance, i.e., the system guarantees a specific amount of CPU cycles between actions, c) are modular and easy to expand with additional components, and d) are rugged and designed for 24/7 operation under harsh environmental conditions, say, high temperatures and humidity, (e) are replaced after many years of continuous operation, mainly because they are constantly connected and interact directly with physical equipment. Actually, from real life observations, it could be also argued that devices even higher levels often rely in deprecated and sometimes running unpatched OS and applications.

*C. ICS Protocols*

The most widespread protocols used in ICS are Modbus, DNP3, IEC-104, IEC 61850, PROFIBUS/PROFINET, EtherNet/IP, OPC, WirelessHART and ZigBee. These protocols were specifically designed to deal with the complexity and the special requirements of the ICS. The operations inside ICS are real-time (deterministic), reliable, safety critical, ruggedized, and sometimes remote. In the most typical cases, the use of serial buses is widespread and the protocols that are used in levels 0-1 are referred to as Fieldbus protocols. However, nowadays, there are protocols that utilize directly Ethernet or TCP/IP stacks depending on the particular use-case. Many of the traditional serial protocols, including Modbus and PROFIBUS, have a corresponding TCP/IP variant, in this case, Modbus TCP and PROFINET, while others, like EtherNet/IP and ZigBee, were designed to work directly over Ethernet and TCP/IP.

It is not the intention of this work to explain each of the above-mentioned protocol in depth [7], [6], but only focus on its security aspects. To this end, by referring to well-studied, real-life major incidents, Section V details on how these protocols can sometimes be exploited by attackers. Moreover, security shortcomings and vulnerabilities of the protocols as identified by the so-far published academic work are given in Section VI.

*D. ICS Security*

Security practices in ICS can be either mandated by regulatory bodies, such as North American Electric Reliability Corporation (NERC) or recommended by entities such as the U.S. Cybersecurity and Infrastructure Security Agency (CISA) and the U.S National Institute of Standards and Technology (NIST). Common key "best practices" are as follows [8]:

- Identify the critical assets that need to be protected.
- Separate the systems into logical and functional groups.
- Implement access control into and between each group.
- Monitor activities.
- Implement a defense-in-depth strategy.
- Limit the actions that can be executed within and between groups.

When first commissioned, many, if not all, of the ICS were kept isolated from other systems, forming a separate OT network. However, due to the rapid convergence of OT and IT, ICS are drawing more attention from adversaries who, as further explain in Section IV, aim either at financial gain, espionage, or sabotage through process disruption, or physical destruction. In addition, many of the aforementioned devices and protocols used in ICS lack security features vis-à-vis their IT counterparts. They instead were created with a focus on assuring the timeliness and availability of the data used for monitoring and controlling critical industrial processes, rather than preserving security services like authentication, data confidentiality and integrity. Often, there are certain misconceptions that peripheral network security measures, including firewalls and "air gaps" can protect from all sorts of cyber threats. Sections V, VI, and VII further stress on the fact that such security measures cannot be conceived as a "silver-bullet" defense for the ICS.

Furthermore, ICS security requires a deep knowledge of the system's specific operations. In some industries, typical operations are the combination and mix of chemicals, refining of oil, and the generation, transmission, and distribution of electricity or energy in general. These processes are usually automated and called control loops in the industrial terminology. Consider, for example, a tank filled with liquid chemicals that get mixed. The level and composition of the liquid in the tank is indicated by sensors. When the combination of these chemicals reaches a specific density, the liquid is removed from the tank using pumps, and more chemicals are poured back into the tank. This shows that control loops are essential to be well-understood given that: a) the implemented security measures should not disrupt the process in any way, and b) ICS and critical infrastructures (CI) are an attractive target especially for competent and well-equipped adversaries.

*E. Critical Infrastructure*

A CI can be defined as the physical and cyber systems and assets that are essential for the uninterrupted functioning of a nation's society and economy. According to the U.S. CISA, there are 16 CI sectors [9], namely chemical, energy, nuclear reactors, materials and waste, water and wastewater systems, healthcare and public health, transportation systems, financial services, critical manufacturing, dams, commercial facilities, communications, emergency services, defense industrial base, food and agriculture, government facilities, and information technology.

Besides the U.S., similar critical sectors have been identified by the European Union (EU) and individual countries around the globe. For acquiring more information on this matter, the interested reader can refer to [10], [11], [12], [13].

Table I: Related Work

| Contribution | Year | Incidents | ICS protocols | ICS Devices | Taxonomy included | Testbeds | ICS Security Framework | Novel Approaches | Estimated Completion Level[†] |
|---|---|---|---|---|---|---|---|---|---|
| [16] | 2012 | • | | | | | | | Minor |
| [3] | 2015 | | | | | • | | | Minor |
| [17] | 2016 | • | | • | | • | • | • | High |
| [18] | 2017 | | • | | | • | | | Minor |
| [19] | 2018 | • | | | | | | | Minor |
| [20] | 2019 | | • | | • | | | | Medium |
| [21] | 2020 | | | | | | | • | Minor |
| [22] | 2020 | • | | | • | | | | Minor |
| [23] | 2020 | | | | | | | | Minor |
| [1] | 2020 | • | • | | | • | | • | Medium |
| This work | 2021 | • | • | • | • | • | (•) | | High |

† This metric indicate the completion of each work; Minor: ◔, Medium: ◑, High: ◕, Complete: ●

The threats against critical infrastructure can be associated with either physical phenomena such as extreme weather, earthquakes, floods, and epidemics or pandemics, or human-related phenomena, including accidents, espionage, acts of terrorism, and cyberattacks. Therefore, the aspects of security and resiliency based on potential threats, are cardinal to the risk management process per CI sector. Since CI is complex, multi-layered, and involves a plethora of stakeholders, more attention is paid to the remediation of risks that are more probable and are also estimated to have a higher impact. This often results in cybersecurity-related measures to be neglected, since other measures, including physical security and the protection from physical phenomena, are often considered of higher priority.

Furthermore, cooperation and communication via information sharing, namely cyberthreat intelligence, is essential, not only in an inter-CI fashion, but also across different CI sectors, as many of them are obviously interdependent. From a cybersecurity viewpoint, CISA [14] and the Information Sharing and Analysis Centers (ISACs) such as E-ISAC [15] are well-known organizations that promote and assist this effort.

## III. Related Work

ICS, CI, and their security issues and challenges have been investigated for several years. However, no particular focus has been given on the technical aspects and the root causes of the described incidents.

The rest of this section elaborates on the most pertinent surveys on this topic and discusses the key differences between them and the work at hand. Our study spans 5 years, i.e., from 2012 to 2020, and the various works are presented chronologically, from the most current to the oldest. Nevertheless, for the sake of completion, and as summarized in Table I, we do provide references to either significant but outdated work [16], or others devoted to more specific areas of ICS [3]. We choose the categories based on the information gathered from the related work and the information we provide in this work.

In [17] McLaughlin et al. survey the ICS cybersecurity landscape and discuss both offensive and defensive mechanisms for various levels of the ICS, including hardware, firmware, software, network and process. The authors focus on vulnerability assessment methodologies, ICS testbeds, attack vectors, say, payload construction and false data injection, vulnerability remedies and a number of secure control architectures. However, the paper does not offer a full-scale analysis of vulnerabilities and real-life incidents in ICS and CI.

The work of Xu et al. [18] reviews the vulnerabilities of common ICS protocols and elaborate on relevant attacks. On top of it, the authors detail on proposed countermeasures and current testbed implementations that can be used to perform both offensive and defensive research. Nevertheless, their work completely neglects wireless protocols exploited in the context of IIoT.

Hemsley and Fisher [19] present a study of publicized security incidents against various CI sectors, and elaborate on the diverse types of adversaries. Similar to our work, they focus on the most significant incidents trying to provide a complete view of the vulnerable components per type of CI. However, several of the incidents included in their work lack a detailed analysis. Furthermore, no discussion is made the impact of vulnerabilities that affect specific ICS protocols and devices.

Volkova et al. [20] survey the several ICS communication protocols, namely Modbus, OPC-UA, TASE.2, DNP3, IEC 60870-5-101, IEC 60870-5-104, and IEC 61850. Some of these protocols, along with their vulnerabilities, are also discussed in this paper. The authors categorize the various protocols based on whether they cater for confidentiality, integrity, and availability. Potential security breaches in control system communication protocols based on the vulnerable protocols along with real-case scenarios and security recommendations are also put forward. Pliatsos et al. [1] discuss the security of SCADA communication protocols. Additionally, they present a number of security incidents against ICS and CI along with relevant objectives and threats. They examine various proposals that aim at enhancing the security of SCADA systems in terms of attack detection. Moreover, they detail on the most common attack types against SCADA systems and offer an extensive presentation of SCADA security testbeds. On the downside, both the above-mentioned works do not provide a fully-fledged analysis on ICS security incidents from a technical viewpoint and lack of an analysis of wireless protocols used in ICS.

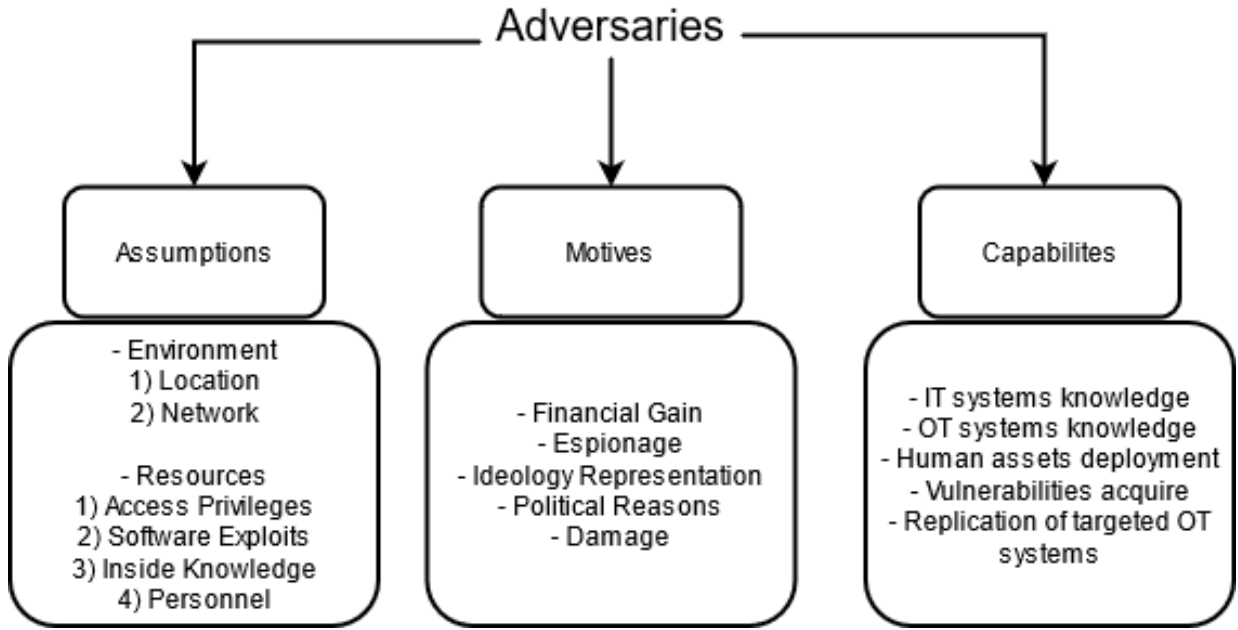

Figure 3: The three main components of an adversary model in ICS.

The work from Bhamare et al. [21] discusses the general state of play of cybersecurity in ICS. The key topics presented in this work are the integration of ICS with cloud-based environments, and the use of machine learning techniques in aid of ICS cybersecurity. Moreover, a thorough categorization of approaches for ICS cybersecurity is offered. However, neither a detailed investigation of major real-life incidents nor the specific vulnerabilities pertaining to ICS equipment and protocols are presented.

Ahmadian et al. [22] perform a survey around cybersecurity incidents in ICS. They group these incidents into the attack and non-attack related based on specific characteristics that govern them. They present information about the diverse attack sources, the entry points that may leave room for realising such an incident, as well as its direct impact. The authors abstractly analyze some of the considered incidents, without however presenting adequate technical details that would allow the reader to grasp the precise nature of these events. The work from Alladi et al. [23] details on attacks against ICS and CI. Yet, the technical details provided are once more limited, and no discussion about ICS protocols and relevant devices is included.

Given the above discussion, the current work not only contributes an extensive and state-of-the-art analysis of the various major security incidents against ICS and CI but also puts forward practical and theoretical security shortcomings that specifically affect diverse operational layers of the Purdue model. In this respect, the work at hand is not only full-featured, thus complementing the hitherto literature on the specific subject, but also is anticipated to provide a spherical view regarding the ICS and CI security state of affairs.

## IV. ADVERSARIAL MODEL

Adversaries are individuals, groups, or organisations who attempt to compromise the security of CI, and possibly disrupt its operation. This section elaborates on their types, namely outsiders, insiders, industrial espionage actors, terrorists, and nation-state actors, and discusses their characteristics and motives. Specifically, as shown in Figure 3, the segregation between the diverse types of adversaries is made on four factors; resources, location, motives, and skill level.

- **Outsiders**: They are the most common adversaries in IT and OT environments. They exist outside the physical and network locations of the ICS environment. Depending on their resources and skills, they may possess prior knowledge of the ICS-specific assets. They try to acquire the desired knowledge that will allow them to penetrate the ICS network. Depending on their skills, the adversaries can map and/or access IT and/or OT systems. They can also collaborate with someone on the inside of the organization to acquire that information or even gain access to the targeted assets.

- **Insiders**: Malicious insiders can cause harm to the systems by leveraging the information they retrieve based on their duties. Often, these are disgruntled employees with access both to the facility and the network. There are also non-malicious insiders that can bypass security policies that unknowingly lessens the security of the system. Mainly they leverage their knowledge and level of access, as tools to perform their actions. Depending on the information about the ICS environment and the type of software that runs in various systems, their actions can have different impact e.g., production downtime, catastrophic damages.

- **Criminals / Hacktivists / Script kiddies**: These are actors that usually perform their actions for financial gain or hacktivism. We can categorize them as outsiders. They use common and sometimes widely available tools that are not drastically modified and can be easily detected with the properly deployed countermeasures. Although their capabilities are limited in terms of resources and knowledge of the process that runs in ICS and CI, they can cause damage by exploiting common systems such as MS Windows PCs and servers.

- **Industrial espionage actors**: Industrial espionage has as a primary goal the exfiltration of information about the inner workings of ICS and CI. This includes stealing customers' personal information or acquiring confidential data related to some product, say, a vaccine. Especially with the arrival of IIoT the collected big data can provide crucial information to this type of adversaries. Usually, these actors can be categorized as outsiders, e.g. business competitors, either foreign or domestic, who have some knowledge of the target's premises. These actors have skills that provide them with the capability to acquire a great amount of information (such as screenshots, blueprints, application logic) and often collaborate with insiders, but at the same time, they wish to remain as stealthy as possible.

- **Cyber-terrorists**: This category also includes extremists, hacktivists, and other organized cyber-criminals. They target ICS and CI with the purpose of creating havoc and possibly spreading their ideology. For that reason, they often use tactics to make the public pay attention. They are categorized as outsiders. They are familiar with the physical premises of the targeted ICS and CI, and they persistently attempt to gain access to the network. This type of adversaries can be supported by foreign states and other interested domestic or foreign entities. This type of actors may additionally have in their payroll individuals with adequate skills and tools, and their strength depends on the resources, both monetary and non-monetary. They acquire tools from various resources, which are not widely accessible to other actors. The use of the human element - through the use of social engineering techniques - to deliver or initiate the exploits, is another characteristic of their tactics.

- **Nation-state actors**: They are considered the most powerful, well-equipped, and skilled outsiders. Having, by definition extensive and sometimes unrestricted resources, they can target and damage a diverse set of CI. Through persistent reconnaissance, collusion with insiders, and the use of social engineering techniques, they can acquire extensive knowledge of the physical location and partially knowledge of the ICS intranet. They can also leverage the tactics and techniques of industrial espionage. The attacks that originate from this type of actors can be performed as a means to test their capabilities, apply pressure to other nations or organizations for political reasons, polarize public opinion on controversial or other key matters, and cause shock, confusion, disorder and even harm to the administration and citizens. Their tactics are often performed under high secrecy with the ultimate goal to maintain a foothold in the targeted network. Their arsenal comprises a mix of legacy and specially crafted, highly-sophisticated tools that might also include zero-day exploits. They have the capability to replicate the OT network, partially or in its entirety. Under the prism of modern cyberwarfare, it should not be neglected that cyberattacks against CI is a powerful arrow in the quiver of hybrid threats.

We should also mention that based on the analysis of Caltagirone et al. [24], often there is not only one actor, but rather an activity group that usually operates in a specific geographical area, verticals or mission. These adversarial groups can be combined or split based their motivations and intent.

## V. INDUSTRIAL CONTROL SYSTEMS AND CRITICAL INFRASTRUCTURE INCIDENTS

This section details some of the most well-known incidents that target ICS and/or CI. The description of each incident is conceptually split into six parts/axes, namely, infection, spreading, payload effects, C&C (if any), variants, and key factors that enabled the attack. The chronologically arranged Figure 4 summarizes the relevant information. The selection of the specific incidents is based on the fact that the pieces of information collected were adequate to provide a complete view driven by the above axes.

### A. Stuxnet

Stuxnet [25] is considered the first malware specifically designed to inflict damage against equipment residing in an ICS. The malware is also supplemented with industrial espionage capabilities. Evidence indicates that the malware's primary target was the Natanz nuclear enrichment plant in Iran.

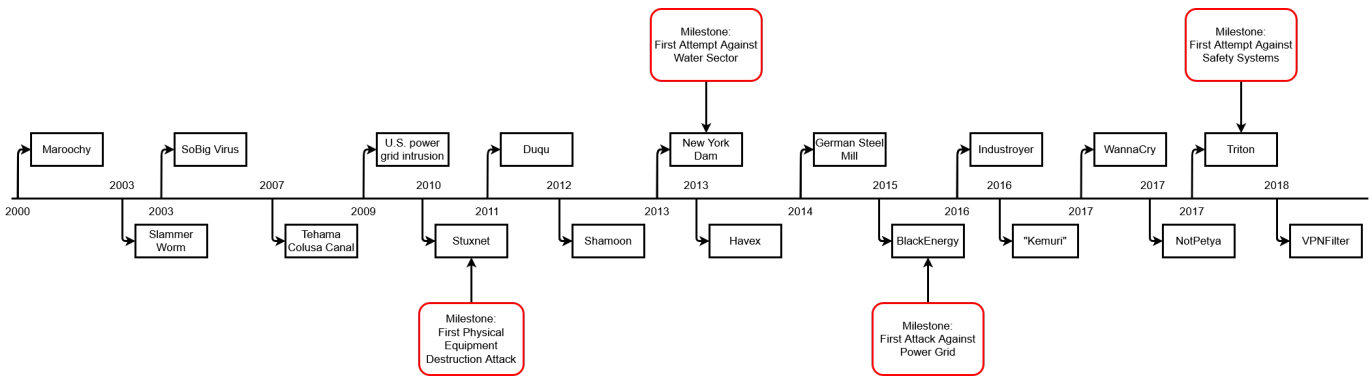

Figure 4: Timeline of the discussed incidents

Stuxnet delivers its first few infections via physical means, i.e., via a USB flash drive. From that point on it attempts to spread to other workstations in the target network via multitude of alternative zero-day vulnerabilities, including a) USB flash drives (CVE-2010-2568), b) the Windows Print Spooler service (CVE-2010-2729), c) network shares or the Server Service (CVE-2008-4250), d) local privilege escalation (CVE-2010-2743) and e) WinCC and PCS 7 SCADA system (CVE-2010-2772). Interestingly, it is programmed to only infect up to three victims, and then it erases itself from the infected media. Moreover, after a specific, hardcoded date (in the discovered cases, June 24, 2012), it ceases any infection attempts.

On a second stage, Stuxnet redirects its focus on spreading to "Field PGs", the specific type of EW made by Siemens which among others are used to program the target PLCs. To this end, the malware infects the WinCC, a Siemens software designed to monitor and write data to the PLCs. The malware takes advantage of hardcoded credentials embedded in the software.

As soon as an infected Field PGs connects to a Siemens S7-315 PLC for programming it, the malicious payload gets uploaded. The payload itself alters the control logic of the PLCs. The payload includes a rootkit destined to hide all the malicious actions performed. In the case of a successful PLC infection, Stuxnet monitors the PROFIBUS connections for 13 days [25]; this was the timeline where the uranium was added to the centrifuges. Then, it alters the operational speeds of two frequency converter drives, manufactured specifically by the Finnish Vacon or Iranian Fararo Paya. Firstly, all functions related to the graceful shutdown of the system in case of a malfunction are disabled. In parallel, a sequence of actions that affect the centrifuges is performed with a duration of 27 days. Initially, the malware records the benign process events to find the current operating frequency and then increases the rotating frequency to 1410Hz for 15 minutes. Then, normal operations are resumed for 27 more days. On the subsequent cycle, it forces the frequency to rapidly drop to 2Hz, followed by an extremely rapid increase to 1064Hz. The changes in rotating frequency make the pressure applied to the inner walls of the centrifuge [26] to get increased and damage the container levels. It is believed that the malware aims to possibly to accelerate the degradation rate of the equipment, which in turn leads to higher operational costs due to more frequent equipment replacement.

The malicious code includes functionality that allows the attackers to control and update Stuxnet through a C&C server. However, since the Field PGs is expected to operate in an isolated network, the malware does not aim in conventional, direct ways of communication, e.g., the Internet. Rather, it aims to compromise and then use the external contractor companies as proxies. Commands and updates are first pushed to these naturally less secure networks that they would eventually penetrate the siloed networks of the facility via conventional means, e.g., USB flash drives. The C&C functionality is implemented to report the infections to two domains, namely *www.mypremierfutbol.com* and *www.todaysfutbol.com*, to inform the attackers whether the Siemens Step 7 programming software is present.

An alternative, earlier version of Stuxnet (version 0.5) was found and analyzed in 2013 [27]. The particular version includes no Windows-based exploits, but rather it replicates only through Step 7 files by utilizing an arbitrary code execution vulnerability (CVE- 2012-3015). This version also introduced the "kill switch" module that prohibits the execution of any malicious operation after July 4, 2009. The main difference of Stuxnet 0.5 is that it aims to control the centrifuge valves that are handled by an S7-417 PLC, instead of the frequency converter drives. More specifically, it monitors the pressure inside the centrifuge via the infected PLC, and as soon the pressure reaches a specific level, it closes the valves.

Stuxnet was probably designed taking into account the detailed information about the specifics of the target environment. Mainly, to increase the probability of penetration to the target environment, the malware relies on a multitude of zero-days. The manipulation of the I/O process image, as it has both read and write capabilities, is used to intercept the benign values of the output and ensures that are not written to the process image output, therefore deceiving the operators. This was a common design flaw in ICS and can be easily exploited, as indicated by Langner [28]. Finally, the compromise of the digital certificates indicates a powerful adversary with high determination, capabilities and resources behind this attack.

## B. Duqu

Duqu [29] is a malware discovered in Hungary by the Laboratory of Cryptography and System Security (CrySyS Lab), that shares behavioral similarities with Stuxnet. For instance, the malware hinges on compromised digital certificates that are used to sign device drivers and exploits zero days as part of its offensive repertoire. Yet, unlike Stuxnet, Duqu's main purpose lies only in cyber-espionage, i.e., the leakage of valuable information from ICS and critical infrastructures. There is no publicly available information regarding the organizations that were impacted, although the malware samples analyzed by Symantec were obtained from ICS entities [30].

Existing studies indicate that a zero-day vulnerability in Windows TrueType font (CVE-2011-3402) is responsible for the initial infection of the target machine. This vulnerability enables remote code execution (RCE), allowing the attacker to perform arbitrary commands through the Internet. Presumably, the malware reaches the target host and exploits that vulnerability via MS Word documents. Then a driver, i.e., an *.sys* file, is used to inject the malicious payload at system boot. This driver can be either singed with a compromised digital certificate or be unsigned. The actual payload is encrypted, and its decryption happens after the driver's initialization and after the malware has verified that the Safe and Debug modes on Windows OS are disabled. The payload is a *.dll* file masqueraded as a *.pnf* file (the particular filetype contains setup information and facilitates the installation process of programs in Windows), that is loaded into the *services.exe* system process. Some of the Duqu's *.dll* functions match those found in Stuxnet, which indicates an association.

After the original infection, a different methodology is used to disseminate Duqu to more hosts in the network. More specifically, (a) Duqu downloads a keylogger from the C&C server and uses it to steal the administrator's credentials of critical servers in the targeted network, and (b) alternatively, Duqu gets copied in corporate shared storage folders. To automatically run, it assigns itself as a remote executable task in the MS Windows Task Scheduler service.

The malware spawns a Remote Procedure Call (RPC) component to communicate with an external malicious C&C server from within the compromised network. This allows the malware to transfer stolen information, and to receive commands or even download upgrades that extend its capabilities. To remain stealthy, it employs a number of external compromised servers as communication proxies. Interestingly, in some of the real-life incidents, the servers appeared to have been compromised simply by brute-forcing the root credentials [31]. For hosts unable to directly communicate with the C&C, i.e., those deprived of Internet access, Duqu is equipped with P2P communication capabilities.

After the infection phase, as a first step, the malware attempts to bypass antivirus programs installed in the system. As a second step, the payload gets downloaded from the C&C server. It contains two *.dll* files aiming to steal data, namely, by recording keystrokes, taking screenshots, enumerating files from all drives, and storing them into temporary locations in the system after compressing and encrypting them. Finally, the captured information is later transmitted to the C&C.

In 2015 Kaspersky detected the variant that they coined as Duqu 2.0 [32]. This newer version also leverages an additional zero-day vulnerability (CVE-2015-2360). It is also a cyber-espionage tool that is very modular and aims to the extensive collection of system and user information. Moreover, it is believed that Duqu 2.0 targeted the P5+1 events regarding the negotiations of Iran's nuclear program, cybersecurity, and telecommunications companies. The latter was targeted in an effort to provide the attacking group with more capabilities, and information [33].

Duqu was able to perform all its actions due to the use of a zero-day vulnerability and social engineering techniques that deceit the employees to download the malicious attachment and initiate the attack. In addition, similarly to Stuxnet, the adversaries demonstrate access to considerable resources and high determination, e.g. the capacity to compromise the digital certificates.

## C. Shamoon

Shamoon [34] is a malware that aims in rendering the computers inside target organizations unusable by wiping their hard drives. Among the well-known victims of the malware are the Saudi Arabian oil companies Saudi Aramco and RasGas in 2012. In the case of the former, it is believed that 30K to 55K hosts inside the company's business network were infected, resulting in downtime that span 10 days. Although the ICS network was not directly affected, this incident is an example of how the demise of the IT network may cause an indirect disturbance in OT activities.

Originally, it was believed that the infection took place using a phishing attack. However, further investigation indicates that potentially an insider with access both to the internal network and the Internet might have been involved [35], [36]. As soon as the malware is installed, its dropper component disables the User Account Control (UAC) in the Windows registry. Then, it creates either a persistent service with name "ntssrv" or a scheduled task that executes the payload at a specific time [37]. Subsequently, as detailed in the following, two more components are unpacked, the "reporter" and the "wiper".

The malware spreads to other hosts in the network via network shares [38], installs the wiper module and waits for instructions from the C&C server [39].

The wiper module that is responsible for the files deletion, may attach itself to a standard Windows process in an attempt to better masquerade itself [39]. The wiper includes a signed driver extracted from a third-party disk utility developed by Eldos, namely the Rawdisk [40]. Normally, the installation of disk drivers as well as raw disk access requires administrator privileges. Yet, by first disabling the UAC, and by using special access features provided by the utility, modification to the disk can be performed even in user-mode. Once the wiper module executes, it enumerates all files and appends their names into *.inf* files. These files are then filled with fragments of a *.jpeg* image. The disk overwriting is performed recursively, which may corrupt the Master Boot Record (MBR). Another interesting point is that additional actions, namely, encryption of files, are also supported [37].

The last standard module of the malware, namely the reporter, forwards details regarding the number of deleted files per targeted host back to the C&C server. Interestingly, this module also includes modes for receiving new executable files from the C&C. However, due to coding errors, this module was not functional.

Newer versions of Shamoon were observed in 2016 [41] and in 2018 [42]. The identified 2016 version, includes minor modifications, i.e., a different *.jpeg* image is used and targets different corporations. The 2018 version also includes an updated wiper component, where the *Trojan.Filerase* malware is utilized instead, to perform a deep erase of the disks, rendering them non-recoverable even with the use of forensics techniques. Furthermore, in this version, a list of selected hosts is included a priori, with these targets probably having been identified as vulnerable at an earlier reconnaissance stage.

The malware is able to perform its actions due to the interconnectivity of all computers in the business network, stolen credentials, and the use of a legitimate driver. Since the exchanged data between the IT and OT are used to determine the business's needs and procedures, such catastrophic attacks to the IT network may deprive the ICS of the high-level site operations that support the production process in the OT.

### D. Havex

Havex [43] (also known as Backdoor.Oldrea) is a backdoor malware used by the Dragonfly group to perform espionage against CI mainly in Europe and the U.S. It is also observed that Havex is used in conjunction with the *Karagany* malware to expand its capabilities. Well-known targets involve companies in power and pharmaceutical sectors [44]. After Stuxnet, it is the first malware designed to impact critical infrastructures by targeting primarily OT equipment.

The malware infects the target systems using three tactics [45]: (a) phishing campaigns, (b) "watering hole" attacks, and (c) compromised websites. The first infection method is based on delivering the malware through malicious PDF documents, dropped to the victim via email attachments. In the case of "watering hole" attacks, the evil-doers first compromise websites frequently visited by the victim company. In the compromised websites case, an *iframe* is used to force the automatic download of the *LightsOut* exploit kit. The particular kit exploits vulnerabilities of Java or MS Internet Explorer installed on the victim's machine to drop Havex or *Karagany*. Newer versions of the malware rely on the *Hello* exploit kit instead, which enables the attackers to fully fingerprint the user's system. The third tactic replaces legitimate software distributed by the websites of third-party vendors with a trojan as a form of supply chain attack. Examples of such software are VPN clients or PLC drivers [44].

After infecting the victim's system, the malware modifies the *Temp* and *System* folders of the Windows OS along with the system's registry. Havex tries to collect information about the infected system, including available drives, generic files, email addresses, and ICS configuration files. The *Karagany* addition improves upon these features.

The most distinguishable feature of Havex is its ability to discover networked devices connected to well-known PLC-related ports, namely, TCP ports 44818 (Rockwell), 102 (Siemens) and 502 (Schneider Electric). The malware module responsible for this process relies on the Distributed Component Object Model-based (DCOM) OPC technology that is normally used to interconnect equipment from different vendors, along with the Windows networking (WNet) service, which is used to expose networking functions to MS Windows applications (see Section VI-I).

As a first step, the malware collects information about each OPC servers' version, vendor information and bandwidth [43] [46]. Next, the Havex payload enumerates the OPC tags provided by each server. An OPC tag is a structure that contains information about the data transmitted by an OPC server to any component in the OT network, e.g., a PLC. Tags are transmitted unprotected and in human-readable format, and their contents tend to be self-explanatory. Therefore, observing these tags may give the attackers knowledge about the physical processes running on the network. The malware also has the ability to make a distinction between real OPC tags and other tags that are provided by honeypots [47].

Both the data collected from the IT and OT environments are sent to the C&C servers. Using a custom encryption scheme to protect this transmission [47]. Apart from receiving and storing the exfiltrated data, the C&C server can also provide payloads to equip the malware with additional functionality.

Subsequent Havex campaigns exposed by Symantec [48], revealed to utilize alternative methodologies to install back-doors to the infected systems that are installed via Windows Powershell, be well-known "trojanized" Windows applications, and be disguised as Adobe Flash updates.

The malware was able to infect the facilities based on the fact that the spam campaigns that delivered the malicious PDFs were successful, the compromise of websites of interest, and the replacement of legitimate software from the vendors with the malicious ones that the engineers downloaded. In addition, the OPC scanning module was successful due to the improper isolation of the IT and OT networks in the targeted facilities as well as the open nature of OPC.

*E. BlackEnergy/2015 Ukraine powergrid cyberattack*

BlackEnergy (BE) [49] is predominantly a botnet/DDoS tool, which through the years evolved into a malware suite with additional sophisticated features. Particularly the third iteration of the malware (BE3) is of special interest as it was used in an illustrious attack campaign against Ukrainian power distribution companies in 2015, namely "Prykarpattyaoblenergo", "Chernivtsioblenergo", and "Kyivoblenergo". In these major incidents, BE3 was used only during the early stages of the attack to deliver the payload to the targeted networks and enable the attackers to perform malicious actions against the ICS remotely. As a result, power outages started occurring, which affected approximately a population of 225K in the regions of Ivano-Frankivsk, Chernivtsi, and Kiev.

BE is created with a very modular architecture in mind [50]. The most popular version of the malware (i.e., the one that was used in Ukrainian attacks), can be delivered via the use of Word documents embedded with malicious macros [51]. If macros execution is enabled in MS Word, a malicious VBA script enables the payload to the startup folder of the system. Then, the payload initiates a connection with the C&C server. It is worth noting that all communication is done over HTTP and is encrypted via RC4 [52].

Prior to the attacks that caused the power outages, a thorough reconnaissance stage took place, which is believed that it may have lasted up to six months [53]. During this period, the attackers gathered all necessary information regarding both the IT and OT environments. To do so, several external tools specifically designed for credentials theft, network discovery and scan, remote execution, screen capturing, and key logging were used [54]. The captured credentials provided the attackers with access to the ICS network. Moreover, devices which under normal conditions provide power supply redundancy to communication and data servers (UPS), were also discovered and were re-configured so the attackers could disconnect them at will. During the reconnaissance phase, the attackers also installed the KillDisk component in a network share. When executed at a later stage, this component overwrote the MBR, and deleted logs and system events, making any subsequent investigation of the attack much harder [55].

The last stage of the attack took place on Dec. 23 2015. The adversaries exploited two different approaches to wreak havoc. In the first approach, Remote Access Tools (RATs) were used by the attacker to connect to the HMIs. Additionally, the operators were locked out of their workstations, unable to perform any actions. The second approach was more stealthy, as the attackers issued commands directly to the Distribution Management System (DMS) server using the VPN connections. As a result, the attackers were able to access the HMIs, to open the circuit breakers, and to cause power outages to at least 57 substations.

After causing the outage, the adversaries proceeded to additional actions to amplify the inflicted damage: (a) pushed a malicious firmware update to corrupt the Moxa and IRZ Serial-to-Ethernet adapters [56]; in this way, they effectively reduced all monitoring and control capabilities of the operators, (b) the installed KillDisk was executed and wiped the operators' PCs but also, due to poor network configuration, affected the HMIs connected to Remote Terminal Units (RTUs) [49], (c) disabled the UPS from the communications server to cause further confusion to the operators, and (d) to make matters worse, a DoS was performed against the telephonic center.

To restore power, all operations were switched to manual mode [57]. The restoration process required approximately six hours.

As described in detail in [58], the BE malware was also used against numerous critical infrastructure targets in a campaign that took place one and a half year before the described incident. These targets include railway transportation systems, Ukrainian regional government and archives facilities, mining companies and television broadcasters. Variants of BE have also been identified in alternative campaigns against U.S. critical infrastructure sectors [59].

The malware could succeed its goals due to the lack of security awareness on the operators' side, the detailed acquisition information about the equipment used in the facility, the lack of two-factor authentication for the VPN services, the improper configuration of the firewalls, and the deficiency of security mechanisms in the Serial-to-Ethernet adapter devices.

*F. Industroyer/CrashOverride/2016 Ukraine powergrid cyberattack*

Industroyer [60] (or CrashOverride) is a malware that targeted the Ukrainian power grid on the Dec. 2016 attacks. This attack comes just a year after the BE3 attack (see subsection V-E) but it is much more sophisticated in comparison. Similarly to BE3, the malware has a very modular architecture that allows it to directly access ICS equipment, however this time at the transmission substations. During the attacks, it caused power outages that lasted almost one hour, affecting one-fifth of the Kiev region.

A report from Dragos [61] indicates that the intrusion took place during a phishing campaign that occurred a few weeks after the successful 2015 attack. Once the malware is installed in the victim's PC it starts to scan for legitimate credentials of remote access (or VPN) tools that may provide a direct connection to the ICS networks. The adversaries created users with administrator privileges in the access server so they could subsequently access a database server, that acted as a *data historian* [61]. A historian concentrates all the data from the ICS environment to provide information to the business network. By default, data historian should support unidirectional data flow only from the ICS to the IT network. A misconfiguration that allowed bidirectional data flow was exploited by the attackers to gain a foothold to the ICS network.

The attackers leveraged the tool Mimikatz [62] as a way to capture and reuse credentials inside the ICS environment. Subsequently, they accessed multiple hosts and attempted to create a link between servers. *Visual Basic* and *BAT* scripts were used to move masqueraded *.exe* files as *.txt* files and execute PowerShell commands.

The malware then tries to install and start itself as a system service in order to execute the payload components. The components included fall into the following categories [60]: Backdoor (one primary and one alternative), launcher, wiper, port scanner, ICS protocol-specific malicious payload modules (IEC-101, 104, 61850, OPC-DA), and DoS module for Siemens SIPROTEC protective relays.

The primary backdoor communicates with C&C servers via the Tor anonymity network and activates only in a specific hour of a day. The alternative backdoor is a "trojanized" MS Windows Notepad application that once executed it can run a shellcode downloaded from the C&C server; interestingly various "trojanized" Windows applications have also been observed in most recent Havex campaigns. However, as described in [61] these backdoors are not vital, as the attack could have been conducted even without their use. The launcher component includes two timestamps (17th Dec. 2016 and 20th Dec. 2016) that indicate the attack dates. The data wiper module is executed as the final stage of the attack. It alters the Registry Keys by making them point to an empty string path and rewrites the standard file-path that is used by all ABB software.

The IEC-101 malicious component implements serial communication according to the IEC 60870-5 standard. This component controls COM ports that communicate with RTUs, which are connected to physical circuit breakers to modify their status from closed to open. The IEC-104 component is similar to the IEC-101, but utilizes TCP/IP communication (see Section VI-C). The IEC-61850 component probes and enumerates devices that use the protocols under this specific standard. If such a device exists, it requests data using the Manufacturing Message Specification (MMS) protocol and searches if there are any matching tags to these messages such as *CSW*, that will indicate the presence of switches and circuit breakers [60],[63]. The port scanning tool is custom-made, probably to avoid detection.

The Industroyer's OPC-DA payload scans and lists all the OPC servers that are provided by ABB software, and also attempts to change the state of devices connected to these OPC servers. Similar behavior has also been observed by Havex (see Section V-D).

The DoS tool leverages a vulnerability found in the Siemens SIPROTEC protective relays (CVE-2015-5374), that allows an attacker to send hand-crafted packets to the device in port 50,000 rendering it unresponsive. It should be noted that according to Slowik [61], the authors of this component made a mistake in the byte conversion of IP addresses, and since these IPs were hardcoded, this component of the attack did not execute. If this component performs its intended action, the disruption event can be transformed into a potential physical destruction attack [64].

Industroyer was successful because of the attackers' knowledge of the grid operations and network communications, the infection via spear-phishing campaigns, the fundamental lack of security mechanisms for the ICS protocols and the exploitation of the vulnerability that targeted the Siemens SIPROTEC devices.

*G. Triton/Trisis/HatMan*

Triton [65] (also known as Trisis or HatMan) malware is created to interact with Triconex Safety Instrumented System (SIS) controllers (made by Schneider Electric) and more specifically the Triconex 3008 processor. The malware's intention is to disrupt the safety mechanisms of the controllers in the target facility. However, as FireEye later discovered, due to incomplete implementation the malware unintentionally triggered the forceful shutdown of the controllers.

Evidence indicate that the attackers may have gained access to the OT network almost a year before the actual incident. Due to a misconfigured firewall [66], the attackers got foothold to a SIS EW and through it, they delivered the payload to the target controller using a custom-made TRITON attack framework.

The two main components of the Triton module that infected the SIS engineering workstation are: (a) an executable, namely, *trilog.exe*, which aims to deliver the payload, and (b) a *library.zip* file that contains all the libraries required to communicate with the Triconex SIS controllers. The trilog.exe was developed in Python but was compiled using Py2EXE to be able to execute in the SIS EW where a Python environment is not usually installed. The only parameter that *trilog.exe* receives is the IP address of the SIS controller. To target this IP and establish communication, the TriStation protocol had to be reverse-engineered by the attackers [67].

The authors of the malware were counting on that eventually the physical four-position key switch of the SIS controller will be set on PROGRAM mode by the engineers [68]. In this mode, where changes are allowed to be performed to the controller, the trilog.exe was able to deliver the file *inject.bin* to the controller.

The inject.bin exploits a zero-day vulnerability (CVE-2018-7522), to elevate its privileges, add another file and restore expected permissions. When finished, a dummy program (initiated by trilog.exe), overwrites the part of the memory segment on the controller that stores the inject.bin. In practice, part of the malicious OT payload namely, *imain.bin* is uploaded either to firmware or application the controller's memory region by inject.bin. It provides an attacker with full access of read/write/execute functionality to the controller irrespective of the Triconex key switch position [68], [69].

Furthermore, four modules inside the *library.zip* are used to deliver inject.bin and imain.bin via the reverse-engineered TriStation protocol. The module *TsHi* exports functions used for input and code signing while the *TsBase* translates those functions into specific codes and formats the data. The underlying UDP protocol is implemented by *TsLow*, where the appropriate function code is chosen, serialization and send of the payload to the controller is performed. The last module, *TS_cnames.py*, contains all the function and response codes, as well as the key switch and control program states.

The code for the Triton malware was leaked and can be found in a GitHub repository [70].

Technically, the possible malicious outcomes of Triton's capabilities may be: (a) shutdown of the process through operational uncertainty, (b) forcing the SIS controller to an unsafe state by maliciously altering the SIS logic, (c) removing all the fail-safes that exist to prevent damage, thus creating an unsafe physical condition [71]. In the studied real-life example, only shutting down of the controllers was observed possibly due to attack implementation errors.

The overlook of alarms from the anti-malware system, a misconfiguration of a firewall, the hardware key left on PROGRAM mode [66], and the relevant zero-day vulnerability, made it possible for the attacker to gain access to the EW and the SIS controller.

### H. VPNFilter

VPNFilter is a modular malware that incorporates both reconnaissance and destructive features. VPNFilter's scanning and infection activity was first observed in May 2018 by the Cisco's Talos Intelligence Group [72]. According to their analysis, the scans performed by the malware targeted primarily routers and Network Attached Storage (NAS) in more than 100 countries. The binary analysis performed by Talos shows that the MIPS and x86 are the targeted architectures. It is also estimated that the malware infected more than 500K devices worldwide. While VPNFilter contains ICS monitoring capabilities, it may also affect other types of environments.

Default credentials and unpatched vulnerabilities are highly suspected of having contributed to the infection process. During this stage, VPNFilter installs binaries that try to connect to a C&C server to target devices that run the well-known Linux BusyBox. The infection method is very resilient, as it adds the binary to the Linux task scheduler configuration service *crontab* to persist across reboots.

As a first step, the installed malware attempts to retrieve images from Photobucket and *toknowall.com*. These contain the active URLs of the C&C server in their meta-data portion. This is done as a way of obfuscation. If that practice fails, the malware attempts to directly establish a connection to a hardcoded, public IP. What is interesting is that the RC4 implementation that is used to encrypt the communications contains a similar bug to the one observed in the BlackEnergy malware (see Section V-E). Thus, one may assume that there may be a connection between the authors of the two.

At this point, the most notable instruction that can be received from the C&C is the "kill" function that can be triggered via the *dstr* module, which can wipe the device's storage [72].

Two additional modules of VPNFilter include a packet sniffer (*ps*) and a Tor network plugin. The former is used to extract website credentials and log Modbus TCP/IP packets. The Tor plugin is used to communicate with the C&C anonymously. Frequently, this communication involves the downloading of new modules that extend the malware with capabilities such as data exfiltration and device management. Newly discovered modules such as *ssler*, are able to intercept and manipulate all traffic from port 80 [73], [74]. Another module namely *tcpvpn* can establish VPN connections on compromised devices, thus enabling the adversaries to access internal networks that the infected devices participate to.

Guidelines on how the users can remove the malware from their devices were issued by the manufactures of potentially vulnerable devices [75], [76], [77] and the FBI [78]. Common countermeasures are the hard reset, applying patches, and change of the default login credentials of the devices. The FBI also seized the domain that belonged to the VPNFilter's authors [79].

VPNFilter was able to achieve its goals by infecting some of the most critical components of a network, i.e., routers and NAS servers, that many applications use by exploiting unpatched vulnerabilities. Moreover, the use of some of these devices in ICS environments provides leverage to the attacker as there are primarily created for Small Office Home Office (SOHO) use, and not for use inside the ICS.

### I. WannaCry

WannaCry [80], is a cryptoworm-based attack that affects MS Windows computers. The worm encrypts files in the OS and demands Bitcoin as ransom. Some mission-critical organizations that were affected by WannaCry in 2017 include the National Health Service (NHS) of the United Kingdom, the Spanish telecommunications company Telefonica, and the U.S. delivery service, FedEx. In the case of NHS, it is estimated that the cost from the WannaCry damages is £19 million. [81].

The infection stage uses an existing exploit known as *EternalBlue* that can achieve Remote Code Execution. In turn, the exploit is based on a vulnerability of the SMBv1 protocol (CVE-2017-0145), which is specific to MS Windows 7 and Windows Server 2008 and earlier versions. It should be pointed out that Microsoft addressed the vulnerability even before the original WannaCry-based attack in security bulletin MS17-010 the original release date of March 14, 2017. Nevertheless, the majority of the systems remained vulnerable due to negligence in installing the particular patch.

Once the vulnerable computer is infected, another tool namely, DoublePulsar is used to deliver the ransomware part. The malware runs two executables as services and tries to connect to a specific domain [80]. It also spawns a persistent service that re-runs the executable that is responsible for internal and external network scanning in two threads. Both threads, check if computers have the SMB port 445 open. The analysis of Malwarebytes Labs [82] indicates that the wormable part of WannaCry maybe still be effective on computers behind a NAT or proxy and not just Internet-facing computers.

After the end of the scans, the *tasksche.exe* creates several files, such as images, and READ_ME files used to display messages in various languages. It also searches for files with specific extensions in all disk drives, including networks shares and removable drives [83]. Then, it uses a combination of RSA and AES algorithms to encrypt the files and changes their extension to *.WNCRY*. Some variants of the malware also delete every shadow copy volume that exists on the system. An additional process called *@Wanadecryptor@.exe* displays the ransom message on the screen and alters the wallpaper. All messages relevant to *@Wanadecryptor@.exe* are forwarded via the Tor network.

The malware also includes a hardcoded unregistered domain that is checked only during the primary stages of the infection. This domain acts either as a "kill-switch", or as an anti-sandbox technique that evades rudimentary malware detection and dissection procedures. In the original incident, Marcus Hutchins, a security researcher, identified this domain and proceeded to register that domain himself in an effort to better study the malware [84]. Although there was no impact on already affected systems, it made it possible to stop the spreading of the malware. After the domain registration, attackers tried to perform a DDoS attack against it using a variant of the Mirai malware [85]. In subsequent versions of the malware, two more domains were included, but both also became registered quickly by security researchers [86], [87]. The last observed version of WannaCry did not include any "kill-switch" mechanism.

The WannaCry malware has a modular architecture which allows the malware to possibly drop and execute different payloads to its targets. Furthermore, traffic is encrypted through a custom-tailored Transport Layer Security (TLS)-like protocol. Interestingly, a similar technique was used in the attacks against Sony Pictures in 2014 [88].

As we observe, the malware achieved its goals by using a recently disclosed vulnerability and the negligence of applying updates that can prevent the spread of the malware.

### J. NotPetya

NotPetya [89] is a cryptoworm attack against MS Windows based-hosts. NotPetya started spreading in Ukraine one month after the WannaCry attack. It is considered an evolution of the Petya malware that was released in 2016. Among the victims of this malware are numerous Ukrainian ministries, banks and metro systems, the Heritage Valley Health System and logistics-shipping company Maersk. In the case of Maersk, the estimated loss in revenue from the damages was over $200 million [90].

The main delivery mechanism in Ukraine was the tax application system, M.E.Doc. More specifically, a deficiency to the patch update policies of the company, allowed the attackers to compromise the particular servers [91]. According to Cisco's Talos Intelligence Group [92], the attackers identified the SSH credentials of administrator accounts and injected a backdoor into the M.E.Doc's software update mechanism. The backdoor can establish a connection with a proxy, and from there, it enables the downloading of malware or the uploading of information extracted by the victim.

Once the targeted host systems update M.E.Doc, the malware is also delivered. As a next step, NotPetya, drops the files for the ransomware message, the *.dll* file that contains the ransomware, the masqueraded version of the *PsExec* utility (a telnet-replacement for remote execution of processes) along with the tool Mimikatz [62] in order to perform credentials harvesting. Then, the malware decides its next steps based on the antivirus present on the infected system [93], if any. If a Kaspersky antivirus is present, the module will not proceed to encrypting any files on the victim. If one of the Norton or Symantec antiviruses are installed, the included EternalBlue exploit will not be used to spread the malware to other hosts. It also checks the level of privileges that it has in the system to decide whether it is going to use the credentials theft module.

The malware employs three alternative ways for its proliferation: (a) network enumeration to discover any DHCP services that will allow it to scan for the SMB ports 445 and 139 [94], (b) through the SMB copy and execution, leveraging the stolen credentials, (c) via the EternalBlue or EternalRomance exploits with the purpose to launch a shellcode and inject the malware to the target. Targets were also accessed via the NTLM protocol used for authentication against Active Directory [95].

After that, NotPetya launches its encryption capabilities. Precisely, it reads the MBR and installs a custom bootloader in its first sector, adds the Bitcoin wallet address for the ransom, and reboots the machine. Once the machine reboots, the malware encrypts the MTF as well as all the files in the computer using a combination of RSA and AES encryption algorithms.

Moreover the malware proceeds into several anti-forensics actions [96], [94]. Once it executes, it deletes itself, its associated tools and modules from the disk, running only in memory. It then rewrites that part of the disk with zeros. Finally, it deletes all security, setup, system, and application logs.

Kaspersky Labs reported that it was possible that the attackers may be incapable of decrypting the victim's machines. Therefore NotPetya can also be classified as a wiper malware rather than a cryptoworm [97], [96].

NotPetya was able to spread due to the infiltration of the M.E.Doc's update system, the credentials harvesting and the use of unpatched and outdated Windows machines.

### K. Other Incidents

*1) German Steel Mill:* The attack against a German steel facility in 2014 has been reported by Bundesamt für Sicherheit in der Informationstechnik (BSI) in their annual report [98]. The adversaries managed to gain access to the OT network of the steel plant and cause severe damage to furnace equipment. Due to the subtlety of the issue, BSI never disclosed technical details or specifics regarding the attack.

According to the BSI report, the attackers used spear-phishing and social engineering tactics to establish access to the business network. From there, they assumed access to the OT network and were able to connect to individual control systems. From there, they changed the logic of components that prevent systems failures, e.g., the non-controlled shutdown of the furnace. The attackers demonstrated familiarity not only with the systems inside IT and OT environments but also with the steel production process. According to Lee et al. [99], the components that were possibly impacted by this attack were, PLCs, HMIs, SIS controllers, and alarm systems. The analysts also believe that the attackers' goal was to cause intentional damage directly to the steel production process.

*2) Maroochy Water Services:* The 2000 Maroochy Water Services incident [100] was a targeted attack from a former employee having special knowledge of the internal procedures that typically take place in the specific installation. Using special equipment, the attacker had the capability to issue remote commands to the system. They infiltrated the water systems in the Maroochy area in Queensland, Australia, and their actions caused 800K liters of sewage to be emptied into local parks, rivers, and the grounds of the Hyatt Regency hotel.

According to Abrams and Weiss [100], the installed SCADA system consisted of 142 sewage pumping stations, each of which had two monitoring computers. These computers were equipped with PDS Compact 500 radio transmitters that were acting as RTUs/PLCs to receive instructions from the control center. Due to the wide area of the installation, several repeater stations were also deployed to assist the communication.

Initially, the systems lost communication, and the pumps could not perform their normal operations, thus releasing sewage. The contractor company initiated an audit to investigate the root cause of the issue. Despite altering the identifier of a station, the operators noticed that the old identifier was still used in some of the remotely issued commands. Initially, these prevented the remote commands from being executed, but soon the attacker suspected that alteration and initiated a brute force to discover the new identifiers.

In subsequent incidents, the adversary disabled the alarms at four pumping stations. This time, the contractor company, in coordination with the police, suspected that an ex-employ could be behind the attack. Therefore, they physically located them and found a laptop with a stolen program for SCADA reconfiguration installed, along with Motorola M120 two-way radio and PDS control devices. Evidence retrieved from the laptop logs also indicated that commands from the system program run at least 31 times, which matched the behavior observed in the company's logs.

At that time, the radio communications used in SCADA systems lack security features or had improper configuration. Furthermore, there were no security requirements from the contractor, and the incident response procedures were insufficient. Finally, the simplicity of the logging mechanisms also required from the operators to perform time-consuming forensics in order to identify the source of issue.

*3) New York Dam:* The intrusion of the Bowman Dam in Rye, New York occurred in 2013 [101]. The target was a small dam with insignificant reservoir volume to cause large-scale damage. Yet, the demonstrated technical capabilities of attackers are alarming. According to the Department of Justice report, this attack is also linked to a larger attack campaign against various U.S. financial institutions, including Bank of America, JP Morgan Chase, and Wells Fargo [102].

The attackers managed to gain access to the SCADA system via a cellular modem [103]. Six remote access attempts took place between Aug. and Sept. 2013. Information about the water levels, temperature, and status of the sluice gate was also obtained. The sluice gate of the dam was not operational at the time of the attack according to city officials [104]. No additional technical details exist in the public domain.

*4) "Kemuri" Water Company:* In 2016 Verizon performed a security assessment for a water company (simply mentioned with pseudonym "Kemuri") [105]. The assessment took place after the employees became suspicious of an intrusion due to irregular valve and duct behavior. The malicious actors managed to access the SCADA system and influence the PLCs that regulate the water flow as well as the chemicals blended in. Fortunately, an alert system that was already in place notified the operators in a timely manner, and more disastrous consequences were prevented.

During the security audit, Verizon identified Internet-facing applications associated with critical operations. Moreover, the equipment that existed in the OT network was antiquated, thus unable to receive any updates. All the network connections from customers' applications, i.e., a payment portal, and PLCs were going into a single router, namely, an obsolete IBM AS/400 produced in 1988. Moreover, the AS/400 was managed by a single employee that possessed the required knowledge of the system, thus creating a single point of failure.

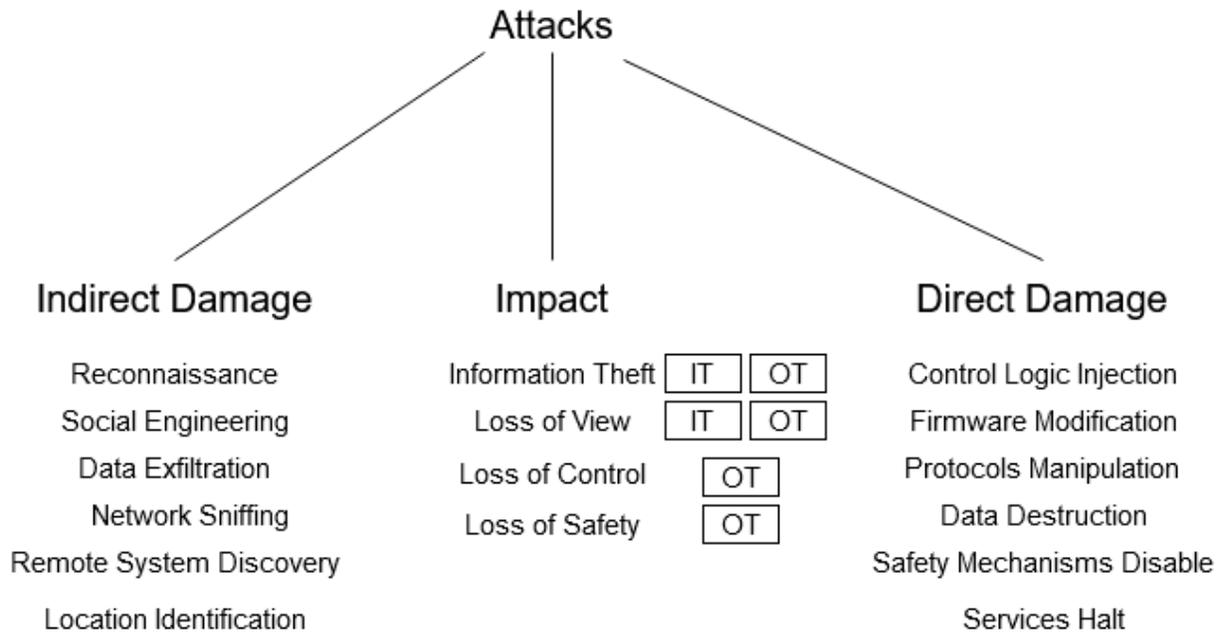

Figure 5: Effects of cyberattacks against ICS

Verizon uncovered unauthorized access to both the business and controls networks. Vericlave [106] stated that an SQL injection attack in combination with social engineering might have been the most probable method of exploitation. From there, the attackers gained access to the Web server that hosted the payment portal and managed to leak 2.5 million customer records.

The attackers retrieved a list of credentials from a configuration file stored in plaintext form in the Web server's filesystem. Interestingly, the credentials were also reused in the SCADA applications. Therefore, they were able to manipulate the industrial process. The logs retrieved from the systems showed that the credentials were used in four separate connections within two months.

The incident occurred mainly because of the inadequate ICS network segmentation, the improper configuration of the services (Internet exposed, access to AS/400 from external IPs), the use of outdated hardware and software, as well as the lack of cybersecurity awareness that could prevent social engineering attacks.

*5) Slammer Worm:* The Slammer worm in 2003 [107] managed to disable the monitoring system of a nuclear power plant in the Ohio Davis-Besse [108]. The worm was based on a vulnerability in the Microsoft SQL Server 2000 (CVE-2002-0649), and it penetrated the nuclear power plant's network via a contractor's laptop that was connected to the business network of the facility. The worm managed to reach the monitoring system by leveraging the improper network isolation and made it inaccessible due to the large amount of traffic that was created. There were not any hazardous physical consequences or data theft from this incident due to the fact the plant was offline for maintenance. Therefore, the impact of this incident in our review is minimal.

*6) SoBig Virus:* In 2003, a shutdown of systems that manage train signals in Florida, U.S. is attributed to the SoBig virus [109]. It infected the SCADA systems via email attachments and propagated quickly. However by this infection, neither major problems were caused in the control process nor data exfiltration. Thus, the contribution of this incident in our discussion is not included.

*7) Tehama Colusa Canal:* A former employee in 2007 [110], installed malicious software on the Tehama Colusa Canal Authority SCADA system that was used to divert water from the Sacramento River and provide various services to the local area. However, no further details for this incident were published.

*8) U.S. power grid intrusion:* Foreign nation-states were reported that they have accessed U.S. power grid utilities in 2009 [111]. The adversaries gathered information about the infrastructure. However, technical details that indicate which systems were compromised and what techniques were used, do not exist in the public domain.

*L. Discussion*

A longitudinal analysis of the attacks against ICS attests that the malware reconnaissance and exploitation phases evolve into much simpler methodologies, but at the same time, they are more severe in terms of the number of impacted systems. As an example of this trend we have incidents like Stuxnet on the one side of the spectrum. Being released in 2009-2010, the malware relies on sophisticated self-propagating functions and zero-day vulnerabilities, but it is custom-tailored to impact specific PLC models. The Maroochy Water Services incident is a case where the attacker has full knowledge of hardware, software and the corresponding configuration for the particular installation. On the other side of the spectrum, later incidents relied on more generic methodologies. For example, the 2015 Ukrainian power grid attack, demonstrated that even without a custom-tailored malware in place and by solely relying on well-known vulnerabilities of the IT and OT systems, an attacker could penetrate the ICS and even impact physical processes causing considerable disruption for a non-negligible amount of time. This tendency has also been observed in more recent incidents that involved the WannaCry and NotPetya malware. We further provide a taxonomy of those attacks based on their damage (direct and indirect) and the impact to the organizations in Figure 5.

*1) Common tools and approaches:* A careful examination also indicates that the adversaries rely increasingly on the use of commodity exploit-kits for the reconnaissance and attack phases rather than developing them from scratch. For example, we see the use of Mimikatz for credentials harvesting by Industroyer and NotPetya. Early incidents like those related to Duqu, leveraged custom keyloggers for that purpose. However, the adversaries abandon these tactics due to the effort and the time that has to be invested in achieving the desired results. Another example is the installation of malware, such as Stuxnet that it uses legitimate drivers. This stealthy approach is also discarded, as the process to acquire such digital certificates, that will sign the drivers, is very laborious.

Actually, a trend observed during the last few years with instances like BlackEnergy, Industroyer and VPNFilter is for malware to adopt a modular architecture that allows extending their attacking repertoire on-the-fly, by relying on existing malware tools.

The use of wiping software, in the attacks conducted by the Shamoon malware, as well as, the 2015 and 2016 Ukraine powergrid attacks, presents a common technique that adversaries use to cover their tracks and to make the recovery of the impacted systems difficult. This is an indicator that well-tested techniques are adopted by numerous actors despite targeting different sectors.

We have also seen that even benign tools that inherently exist in these environments, such as PowerShell (Industroyer, Havex and NotPetya), and OPC (Havex, Industroyer) to be leveraged against the targeted organizations. These organizations should harden, monitor and protect especially their special-purpose tools in order to increase the margin of difficulty of malicious use.

Based on the above information, we assume that the common sources for the tools are: 1) interactive groups such as those leveraged by the nation-state and terrorist actors (see Section IV); 2) black market where the adversaries can obtain exploits that could be we integrated in larger campaigns (e.g. BlackEnergy); 3) common malware kits that exist in the wild (e.g. EternalBlue exploit). However, attribution to specific threat actors cannot be provided with the publicly available information.

*2) Vulnerabilities categorization:* By dissecting the attack methodologies observed in the described incidents, one could categorize the exploited vulnerabilities as: Type 0, which are zero-day vulnerabilities; Type 1, which are known vulnerabilities; Type 2, vulnerabilities stemming from inherently insecure services, protocols and tools; Type 3, vulnerabilities relevant to insecure configuration of networks and equipment; and Type 4, which is social engineering.

Based on the above categorization, we observe that a form of social engineering (**Type 4**) is omnipresent especially during the early reconnaissance stages of the attacks (Duqu, Havex) or as a way of gaining a foothold in the systems of the targeted organizations (BlackEnergy, Industroyer, German Steel Mill, "Kemuri" Water Company). Social engineering attacks can be prevented mainly by educating the employees, frequent training sessions that simulate social engineering attempts, as well as with properly documented policies in place.

**Type 3** vulnerabilities are used by the adversaries to access the OT environment directly (VPN) or indirectly (via the IT systems). This can be achieved with credentials harvesting and/or leverage of misconfigurations. Equipped with this knowledge, the attackers create and test the appropriate payload, and in the end, unleash it to their target. In the Triton incident, the misconfiguration of a firewall allowed the adversaries to gain access to the EW that was used to communicate with the SIS controller. Regular security assessments and the use of multi-factor authentication can mitigate these issues.

**Type 2** assumes that inherent vulnerabilities exist in the protocols adopted. For example, the Industroyer malware issued the commands to the switches and circuit breakers, without the need for any authentication. Once the adversaries reach this level, their tasks become easier (but occasionally time-consuming), even when the asset owners used proprietary protocols. In these cases, the use of modern and updated protocols, that provide better security is needed, although the update process is not trivial in many of the ICS. We expand on this in Section VI.

We also observe that existing unpatched vulnerabilities (**Type 1**) can create devastating effects on the ICS. WannaCry and NotPetya leveraged the negligence of update from the organizations and managed to infect numerous devices. In the "Kemuri" Water Company incident, a SQL injection vulnerability that was unaddressed provided a window of opportunity to the adversaries. The early discovery of those vulnerabilities and the update of the systems when possible, reinforce the security of the ICS.

Table II: Type of Vulnerabilities of the discussed incidents

|  | Type 0 | Type 1 | Type 2 | Type 3 | Type 4 |
|---|---|---|---|---|---|
| Stuxnet ([25]) | • | • |  |  |  |
| Duqu ([29]) | • |  |  |  |  |
| Shamoon ([34]) |  |  |  | • | • |
| Havex ([43]) |  |  | • | • | • |
| BlackEnergy ([49]) |  |  |  | • | • |
| Industroyer ([60]) | • |  | • | • | • |
| Triton ([65]) | • |  | • | • |  |
| VPNFilter ([72]) |  | • | • |  |  |
| WannyCry ([80]) |  | • | • |  |  |
| NotPetya ([89]) |  | • | • | • |  |
| Steel Mill ([98]) |  |  |  |  | • |
| Maroochy ([100]) |  |  | • | • |  |
| NY Dam ([101]) |  |  |  | • |  |
| "Kemuri" ([105]) |  | • |  | • | • |
| Slammer ([107]) |  | • |  | • |  |

[†] Type 0: zero-day vulnerabilities, Type 1: known vulnerabilities, Type 2: insecure protocols, Type 3 : insecure configuration, Type 4: social engineering

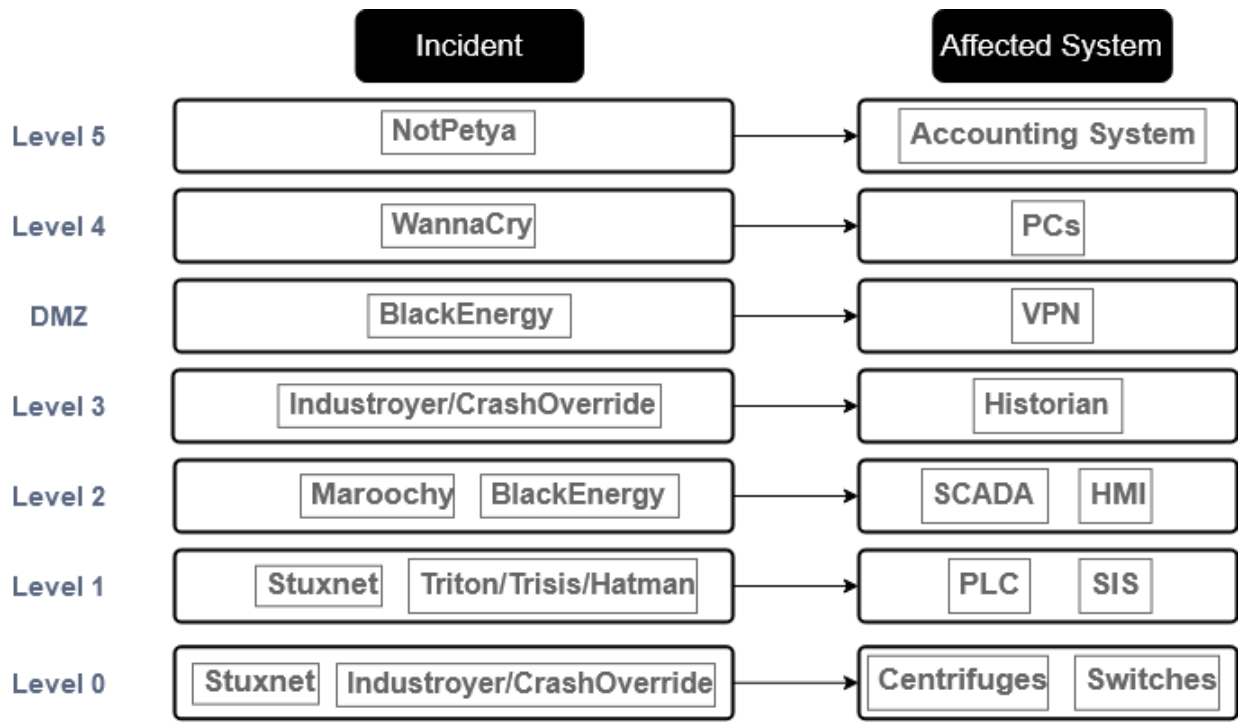

Figure 6: Correspondence of incidents to affected systems; For holistic view see Table III

Zero-day vulnerabilities, i.e., **Type 0** are not common, but dedicated attackers (Stuxnet, Duqu, Industroyer, Triton) can use them against what they consider vital targets. Once again, the earlier those vulnerabilities are discovered by the vendors and the asset owners' systems are patched, the lower are the chances of exploitation. When patching is not possible, due to either system architecture or physical process peculiarities, other countermeasures such as network segregation or use of equipment from different vendors (security-through-diversity) can increase the overall security of the ICS.

To provide a holistic view of the factors that enabled each incident, we categorize the observed vulnerabilities of the discussed incidents in Table II. The reader can see a categorization of each incident and some of the affected Purdue levels in Figure 6. A more detailed one, along with the rest of the related information discussed in this section, is presented in Table III.

*3) Affected Purdue Levels:* Most of the discussed incidents span across multiple levels. For example, Stuxnet infects the PCs and the EWs (levels 4 and 2) that will transfer the project files to the PLC (level 1). While performing its malicious

Table III: All in one taxonomy

| | Type of Vulnerabilities | Affected Purdue Levels | Outcome |
|---|---|---|---|
| Stuxnet ([25]) | Type 0, Type 1 | Level 0, Level 1, Level 2, Level 4 | Economic Loss, Reputation Damage, Societal Damage |
| Duqu ([112]) | Type 0, Type 4 | Level 4 | Economic Loss |
| Shamoon ([34]) | Type 3, Type 4 | Level 4 | Economic Loss, Reputation Damage, Societal Damage |
| Havex ([43]) | Type 2, Type 3, Type 4 | Level 5, Level 3 | Societal Damage |
| BlackEnergy ([49]) | Type 3, Type 4 | Level 4, Level 3, Level 2 | Economic Loss, Reputation Damage, Societal Damage |
| Industroyer ([60]) | Type 0, Type 2, Type 3, Type 4 | Level 4, Level 3, Level 2, Level 1, Level 0 | Economic Loss, Reputation Damage, Societal Damage |
| Triton ([65]) | Type 0, Type 2, Type 3 | Level 2, Level 1 | Reputation Damage |
| VPNFilter ([72]) | Type 1, Type 2 | Level 4, Level 3 | Economic Loss |
| WannyCry ([80]) | Type 1, Type 2 | Level 4 | Economic Loss, Reputation Damage, Societal Damage |
| NotPetya ([89]) | Type 1, Type 2, Type 3 | Level 5, Level 4 | Economic Loss, Reputation Damage, Societal Damage |
| Steel Mill ([98]) | Type 4 | Level 0 | |
| Maroochy ([100]) | Type 2, Type 3 | Level 2, Level 1, Level 0 | Societal Damage |
| NY Dam ([101]) | Type 3 | Level 2 | Reputation Damage |
| "Kemuri" ([105]) | Type 1, Type 3, Type 4 | Level 5, Level 2 | Economic Loss |
| Slammer ([107]) | Type 1, Type 3 | Level 4, Level 2 | Reputation Damage |

Table IV: Mitigation Strategies

| | Secure Remote Access | Patch Management | Credential Management | Network Segmentation | Software Restriction Policies | Outbound Traffic Detection | Execution of Explicitly Allowed Software | Audit Network Hosts for Suspicious Files | Secure Configuration Management | Incident Planning and Response | Awareness and Training |
|---|---|---|---|---|---|---|---|---|---|---|---|
| Stuxnet ([25]) | | | • | | • | • | • | | | | |
| Duqu ([112]) | | | | | • | • | • | | | | • |
| Shamoon ([34]) | | | • | | • | • | | • | | | • |
| Havex ([43]) | | | | • | • | • | • | | | | • |
| BlackEnergy ([49]) | • | | • | | • | • | • | | | | • |
| Industroyer ([60]) | | | | | • | • | • | • | | | • |
| Triton ([65]) | | | | • | • | • | | | • | | • |
| VPNFilter ([72]) | | • | | | • | | | | | | |
| WannyCry ([80]) | | • | • | • | • | • | | • | | | • |
| NotPetya ([89]) | | • | • | | • | • | • | | | | • |
| Steel Mill ([98]) | | | | | | | | | | | • |
| Maroochy ([100]) | • | | | | | | | | | • | |
| NY Dam ([101]) | • | | • | | | | | | | | • |
| "Kemuri" ([105]) | | | | | • | | | | | | • |
| Slammer ([107]) | | • | | | • | | | | • | | • |

1 blank is not applicable or unknown

actions against the centrifuges (level 0), it also affects the view of the operators (level 2). Duqu infects only IT systems however, the information retrieved (such as credentials) can be used in subsequent attacks that target systems in multiple Purdue levels. Shamoon affects IT systems as well, but its wiping capabilities can have repercussions indirectly to the OT side of an organization. Similar behavior is observed in WannaCry and NotPetya. Havex exfiltrates information from the IT systems, and when possible, it also moves to level 4 of the OT environments, when the conditions allow to do so.

BlackEnergy is used to equip the adversaries with the toolset, to perform reconnaissance in the IT and access software such as VPN and remote access tools. Having this type of access, they can connect directly to the OT and perform their malicious actions. The included wiper component can also affect equipment in both IT and OT. Industroyer follows a similar approach. However, it has the additional capability of accessing the equipment that controls switches, circuit breakers at the lowest of the Purdue levels. Triton performs its actions against SIS controllers (level 1) however, it requires prior access to an EW (level 2).

VPNFilter infects routers that exist in numerous Purdue levels, and gather information from the traditionally IT systems (levels 5 and 4) as well as, from routable industrial protocols (level 3). In the German Steel Mill incident, the furnace equipment was damaged, something that requires prior adversarial access to some or all of the previous levels. In the case of the Maroochy Water Services incident, the attacker gained access to the SCADA system (level 2) and issued the commands to the RTUs (level 1), that enabled the pumps to open and release sewage (level 0).

The attackers, in the New York Dam intrusion, accessed the SCADA system in level 2 of the Purdue model and retrieved information about the conditions of the dam. In the "Kemuri" Water Company incident the attackers started by accessing the IT system, and due to misconfigurations, they were able to issue commands to the equipment that regulate the water flow and blends the chemicals via the SCADA system.

*4) Mitigation:* From the described incidents it is evident that, there are inadequate approaches from the targeted ICS organizations to implement the NIST Cybersecurity Framework [113] particularly, in terms of the Identify, Protect, Detect and Respond core functions. Recommendations that are provided by organizations such as the CISA [114], [115] and NIST [8], are also partially implemented or completely neglected in some cases. The incidents examined in this paper could have been prevented or have a lesser impact if strategies such as network segmentation and segregation, application whitelisting (allow-listing), networks monitoring, secure remote access, proper configuration and patch management would have been applied to the targeted ICS as presented in Table IV.

Regarding the patch management, an interesting approach is taken by Spring et al. [116] where the organizations should not take into consideration only the severity of vulnerabilities as identified by their CVSS score [117] but also the unique aspects of each environment and the operations inside it. This can be extremely useful as the peculiarities of each ICS cannot follow the more generic approach taken by traditional IT systems that usually follow the CVSS base score.

Recently, organizations try to focus on how to be more resilient to attacks similar to those discussed in this paper. The limit for resilience is the point where a system cannot perform its desired process. One of the most crucial aspects of resiliency is recovery. For example, incidents that used wiping malware like the 2015 Ukrainian powergrid attack and NotPetya, had severe effects to the recovery process of the affected organizations. Therefore, enhancement of resiliency aspects is very important for all of the organization that employs ICS, to reduce the impact of such adversarial tactics.

We can also conclude that in every incident that had a disruptive impact, the final stage where the attack takes place, is different enough, thus, making the actions of the adversaries unpredictable. The physical process, and the final product can differ significantly even within the same sector (e.g. chemical). Therefore, the adversary must understand how the physical process operates, and employ the appropriate tools in order to maliciously manipulate it. By having a deep understanding of the physical process and various risk indicators, the parties that are responsible for the security of the organization can focus on building security mechanisms around their "crown-jewels" and then move towards assets that can tolerate greater disturbance.

## VI. ICS PROTOCOLS VULNERABILITIES

This section discusses known vulnerabilities of well-established ICS protocols. Such protocols can reside in various layers of the Purdue model, but typically, they follow the architecture shown in Figure 7. More specifically, Modbus, DNP3, PROFINET, EtherNet/IP, and WirelessHART are mainly fieldbus protocols, meaning they are used for communication of industrial devices, including PLCs and IEDs, with components such as sensors, actuators, switches valves, and so on. On the other hand, OPC, IEC-104, IEC-61850, and ZigBee are refereed to as backend protocols. That is, they are used for higher-level interactions of various components in an ICS, such as the communication between a control station and a substation in a power grid. Nevertheless, some of the backend protocols include functionality for fieldbus communication, say, the Manufacturing Message Specification of IEC-61850. Fieldbus communications are time-critical and cost-sensitive, while some of the backend protocols have more tolerance to time delays and can be implemented in more generic hardware. It is worth noting that vulnerabilities may exist in other legacy IT protocols that are also used in ICS, including HTTP, ARP, and Telnet. This section however, intentionally focuses on protocols destined specifically to ICS.

### A. DNP3

Distributed Network Protocol (DNP) [118] is used primarily in the power and water sector. It was devised in 1993 with the goal to address interoperability issues among SCADA components of diverse vendors. It supports both serial and TCP/UDP packet transmission. It is based on standards from the International Electrotechnical Commission (IEC) and its current version, DNP3, has been adopted by IEEE under Std 1815-2010 [119]. Nowadays, DNP3 is an open standard protocol, managed by the DNP3 Users Group, and its latest version was released in 2012. As depicted in Figure 8, the protocol stack of DNP3 comprises three major layers, namely application, pseudorandom, and data link.

Using DNP3, the client, which usually lies in the control station, can perform a polling operation to the server, typically a substation, to receive event data. What distinguishes DNP3 from other similar protocols, is the ability of the server to issue an unsolicited response. Specifically, when a crucial event happens in between the defined communication timeframe, the server can directly notify the clients without delay.

DNP3 has a separate protocol layer called DNP3 Secure Authentication (DNP3-SA), introduced in IEC 62351-5 to provide authentication and message integrity. This layer was added with the intention to secure end-to-end communications for electrical and other types of substations.

The work by East et al. [120] offered a taxonomy of nearly thirty attacks that can be performed against DNP3-oriented systems. The attacks were categorized based on four criteria, namely the target, the threat category, the layers of the protocol that are exploited, and the impact that they cause to the implemented systems. According to the authors, the attacks that can be more harmful are the Rogue Interloper, Length Overflow, Transport Sequence Modification, and Outstation Application Termination, as they can result in loss of control of the inspected system.

In [121] Jin et al. advocated that a SCADA network consisting of DNP3 devices is vulnerable to flooding attacks. The vulnerability stems from the fact that a compromised field device can send a surge of fake unsolicited responses to the control station with the aim of exhausting its memory resources. As illustrated in Figure 9, such an incident would render the control station incapable of storing the values that originate from the legitimate field devices. The root cause of this vulnerability was an error in the vendor's implementation vis-à-vis the DNP3 specification.

Darwish et al. [122] scrutinized the behavior of the DNP3 protocol in smart-grid installations. They exploited some of the vulnerabilities mentioned in [120] and experimented with two attack scenarios, one against unsolicited response messages, and another against data set manipulation. The former blocks unsolicited messages from reaching the outstations, while the latter gives the ability to the assailant to obtain more information about the control station. Both attacks were tested in a simulated environment. These attacks were possible because the protocol does not provide message confidentiality, integrity, authentication, and authorization. The authors also proposed the use of a host-based intrusion detection system (IDS) in each relevant IED to identify such attacks promptly.

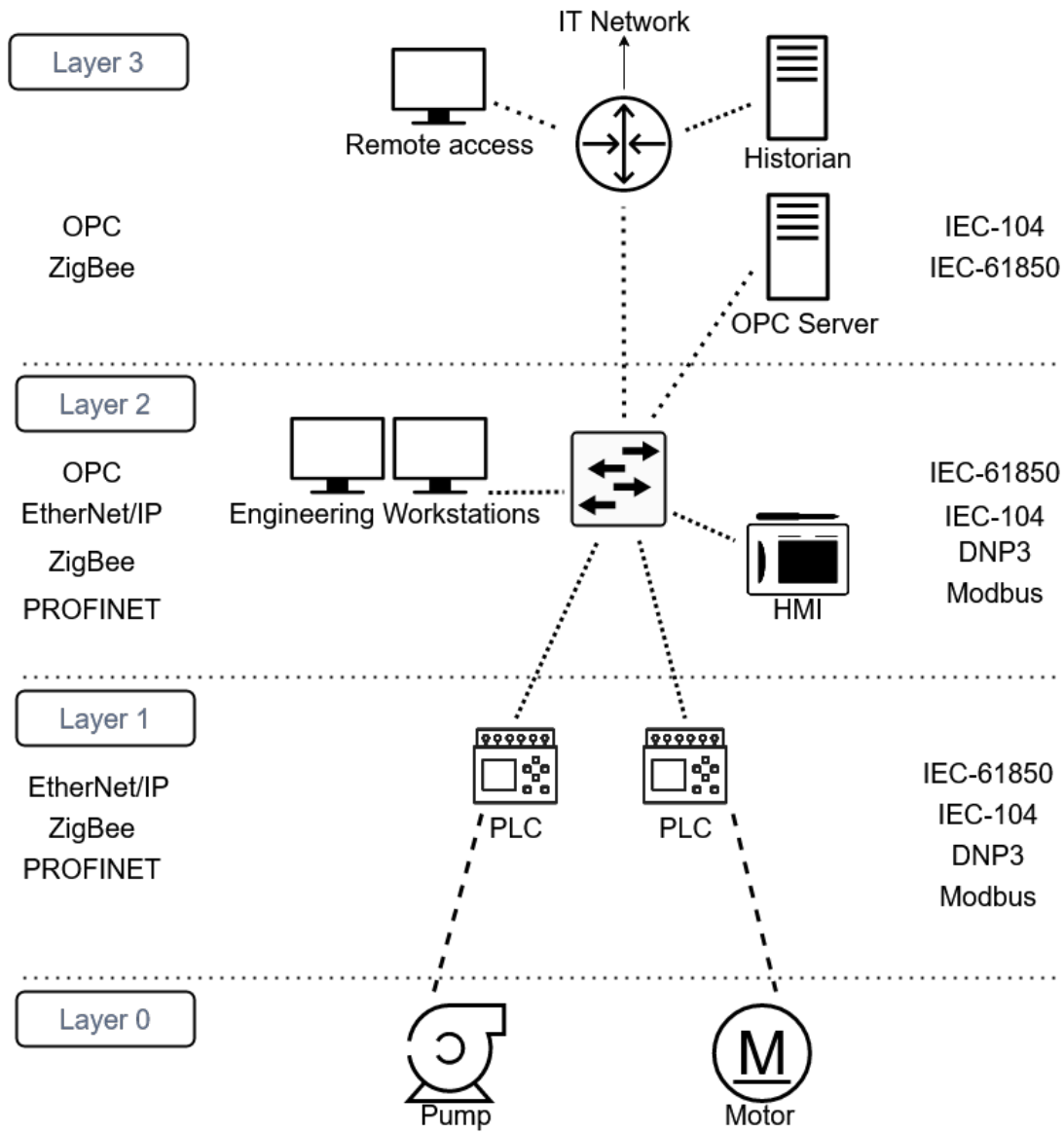

Figure 7: ICS Network Protocols

Another work by Darwish et al. [123] presented an approach that can be used to model DNP3 attacks against the smart-grid realm. Their setup comprised a virtual environment where a MitM attack took place. Particularly, the authors' practical experiments concentrated on sniffing and selectively dropping packets that were exchanged between the client and the server. Additionally, they selectively modified and injected packets to affect the status of the outstation and the control center. On the other hand, the authors' theoretical approach was given as a competition game that can be modelled using game theory. By doing so, they demonstrated the possible combination of strategies for the attacker and the control station, based on one type of MitM attack and its possible outcome. Possible detection and mitigation strategies include the tracking of the network round-trip time and the introduction of a packet pass/drop algorithm.

Rodofile et al. [124] implemented an extension for the Scapy Python library with the purpose of crafting DNP3 packets. Their ultimate goal was to validate some of the attacks presented by East et al. [120]. Specifically, in their experiments, the authors examine eavesdropping, MAC address alteration, frame length overflow, and endpoint device configuration capture type of attacks. Based on their results, only the eavesdropping attack was straightforwardly successful. At the same time, the rest of them worked under specific circumstances, namely address alteration and configuration capture, or evolved into other types of attacks, say, a length overflow developed into a DoS attack.

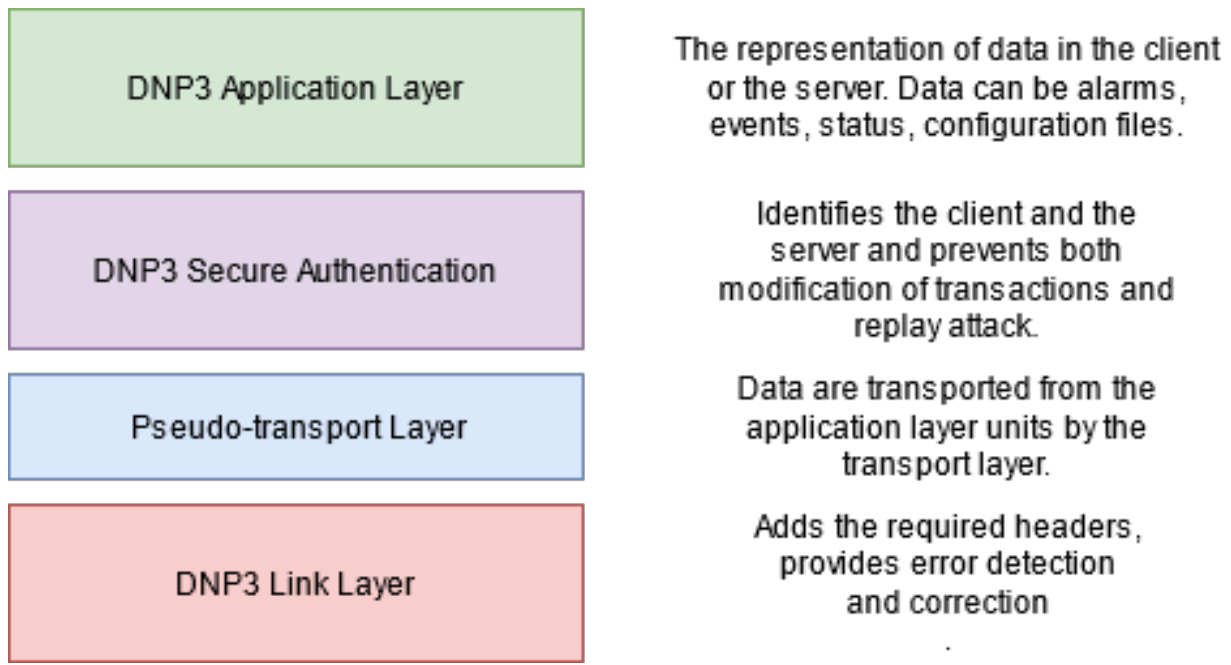

Figure 8: The DNP3 protocol stack

Crain and Bratus [125], demonstrated the use of a fuzzing approach that identifies vulnerabilities in the DNP3 protocol [126]. Their methodology unveiled existing vulnerabilities in various DNP3 implementations, including the lack of message confidentiality, integrity, authentication, and authorization. This comprised the basis of replay attacks as well as, the spoofing of measurement data. The authors also raised concerns regarding (a) the selective use of the authentication mechanisms of function codes in the DNP3-SA, and (b) the ambiguities of the parsing mechanisms in the Aggressive Mode (AGM), a mechanism that provides lower security level compared to the full challenge-response one.

Siddavatam and Kazi [127] examined DNP3 protocol from a security viewpoint and proposed the use of the DNP3-SA framework to achieve resilience against replay and reconnaissance attacks. Once again, these assaults were possible due to the lack of basic security services, namely message confidentiality, integrity, and authentication.

The possibility for unauthorized message modification, replay, and spoofing attacks has been identified in DNP3-SA by Amoah et al.[128]. They stressed that the inefficiency in the authentication property of the protocol could lead to a replay of valid messages in AGM, thus allowing an attacker to execute commands on the stations. Furthermore, the aggressor can write or read values and launch applications at the IED of the compromised station.

Given the above discussion, a key observation is that DPN3 lacks the support of basic security features such as confidently, integrity, availability, and authenticity. Simply put, anyone who is able to reach devices at that level of the ICS, can straightforwardly inspect and manipulate the exchanged messages. Additionally, when solutions like DNP3-SA are introduced, they include features that attempt to balance security and performance, which however, can also be a vulnerable point in the system.

### B. Modbus

Modicon Communication Bus (Modbus) [129] is a commonly used application layer protocol in SCADA domains and the most prolific in industrial manufacturing environments. It was first introduced in 1979 as a serial communications protocol. While originally created by Modicon (now Schneider Electric), Modbus became an open standard in 2004 and is now managed by the Modbus Organization. Nowadays, it includes several variants, including Modbus over TCP/IP, Modbus over UDP, and Modbus Plus (Modbus+). The protocol stacks of Modbus over TCP and Modbus serial (RTU), which are the most common, are depicted in Figure 10. Its current version is 1.1b3, which also includes a Modbus TCP-Security version that works over TLS.

Modbus over TCP utilizes a client/server architecture, where the device acting as a client frequently polls data from one or more devices that operate as servers. A transaction in Modbus begins with the client requesting data from the server using a specific Function Code. If the requested FC is provided by the server, the latter entity will respond with the corresponding data. Otherwise, the server will respond with an Exception Code. Unlike DNP3, Modbus does not support unsolicited responses from the server to the client, thus exposing a lesser attack surface, but loosing functionality regarding the prompt handling of sudden and urgent events.

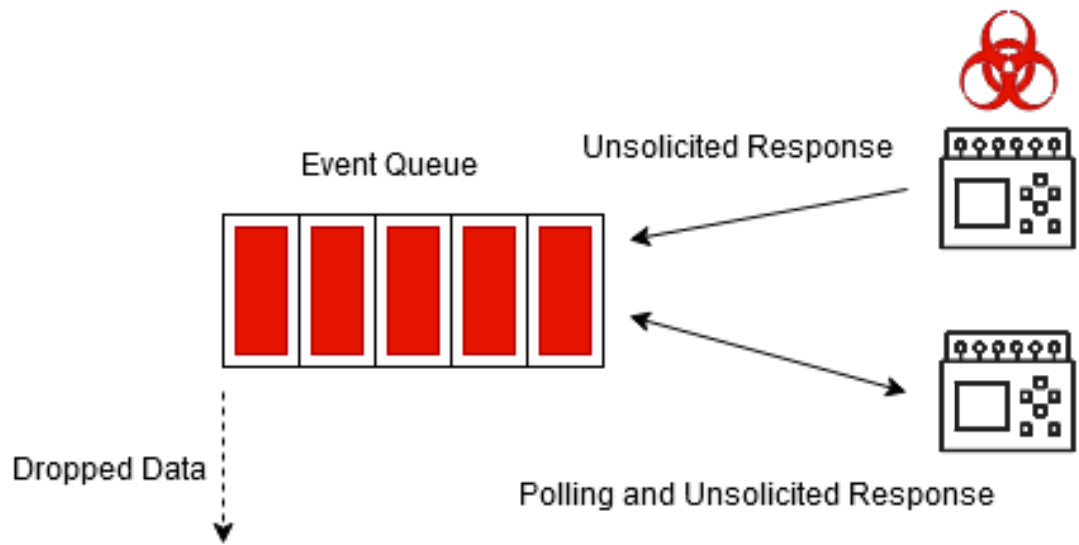

Figure 9: DNP3 buffer flooding attack

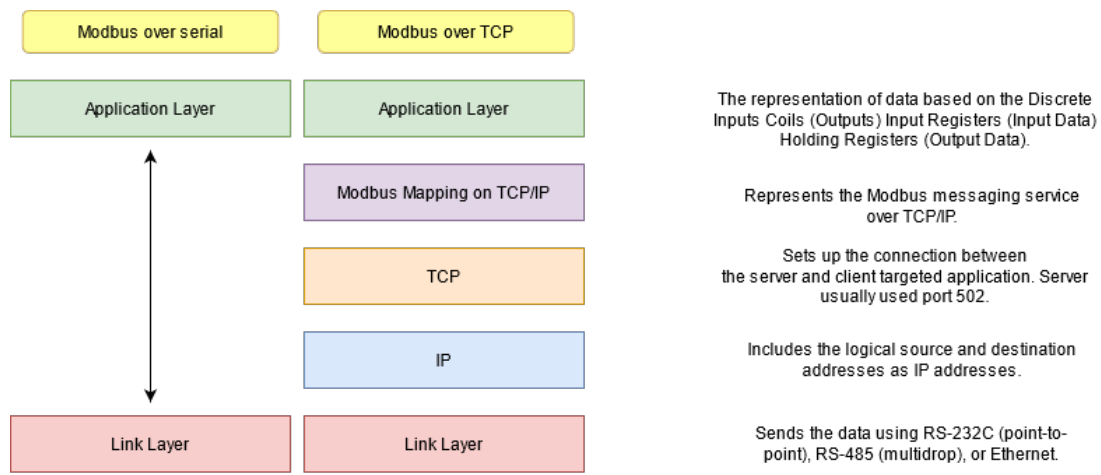

Figure 10: The Modbus protocol stacks

Possible vulnerabilities in the Modbus specification and major implementations of the protocol were investigated by Huitsi [130]. They identified that spoofing, message replay, and flooding attacks could be conducted by leveraging the identified vulnerabilities, which, however, are not mentioned. Finally, they provide a categorization to indicate the criticality of each attack that can be used by the asset owner to protect the affected environments.

Morris et al. [131] detailed theoretical attacks against industrial equipment that rely on Modbus. The authors underlined that due to insufficient security mechanisms for data integrity and availability adopted by the protocol, data injection, and DoS attacks may be feasible. Additionally, the authors aimed at inferring robust intrusion detection rules for the protocol. The work by Gao & Morris [132] described and tested several cyberattacks against systems based on Modbus. Specifically, the authors considered assaults that fall under four umbrellas, namely reconnaissance, response injection, command injection, and DoS. The attacks were implemented and triggered by two control systems. To detect such attacks and ease post-incident analysis, the authors detailed several standalone and state-based IDS rules.

After confirming that Modbus is prone to flooding attacks, Bhatia et al. [133] devised and assessed anomaly and signature-based detection as a means of mitigating them. They conclude that signature-based detection conducted via Snort rules requires a carefully chosen threshold value, while on the other hand, anomaly-based detection is typically unable to identify the attack rapidly.

Nardone et al. [134], formally assessed and evaluated the security of the Modbus protocols in terms of the security features each variant provides. The authors conclude that the main vulnerabilities in Modbus stem from the fact that it fails to provide data integrity, availability, and authentication. Consequently, attackers can perform replay and spoofing attacks by acting as a man-in-the-middle, as illustrated in Figure 11.

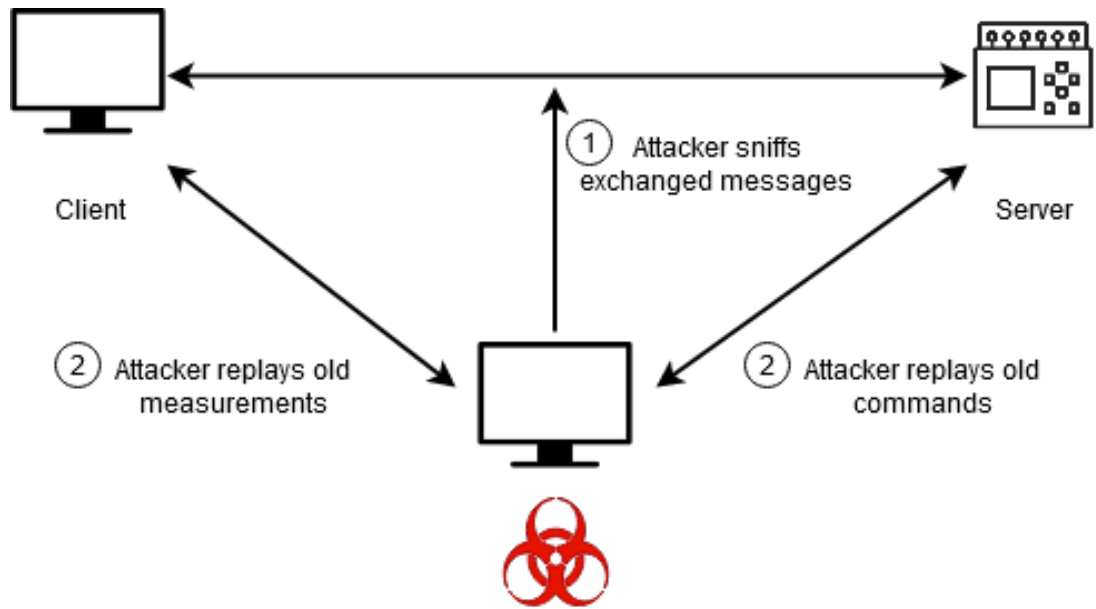

Figure 11: Replay attack on Modbus

The work by Tsalis et al. [135] demonstrated that even in the presence of encryption, side-channel attacks may reveal information on Modbus RTU protocol messages. They specifically show that any web interface that implements unpadded encryption using stream cyphers like RC4 or block cyphers that rely on certain modes, including CFB, GCM, OFB and CTR is prone to differential packet size attacks. Simply put, by default, Modbus functions construct payload in a very similar way which along with the limited number of Modulus Function codes, leads to low entropy in packet sizes, thus uncloaking functionality of the system. To verify their claims, the authors exploited Shodan [136], searching for devices with open Modbus ports over TCP/IP and TLS-protected web traffic.

Using a testbed comprising of virtual machines running on Linux, Parian et al. [137] detailed on two attacks. In the first attack, if the host is infected with malware, the packets could be manipulated before they reach legitimate software. The second attack described is a classic MitM scenario using ARP spoofing that can redirect specific packets to the attackers' machine, where they can be altered. A very similar MitM attack on Modbus has also demonstrated by Chen et al. [138]. In addition, the same authors tested a TCP SYN flood attack with the aim of flooding a Modbus slave with TCP connection requests allegedly originating from clients.

Even if Modbus is one of the oldest and most well-established industrial protocols, it has not until recently included any mechanisms that target the provision of fundamental security services, including the integrity of messages and availability. Efforts like the Modbus/TCP Security protocol, aim to decrease the vulnerable spots in the protocols, by investing in proven solutions such as TLS. However, these enhancements have not penetrated the market yet since the upgrade of ICS components is not trivial and some applications speed is more important than e.g. confidentiality.

### C. IEC-104

IEC 60870-5-104, commonly known as IEC-104, is a set of industrial protocols that belong to the greater family of standards IEC 60870 defined by the International Electrotechnical Commission (IEC) [139]. Formed in 2006, it is an extension of the IEC-101, which operates over serial communication. The standard uses TCP/IP interfaces that can be used for both LAN and WAN communications. The interoperability list pertaining to the various characteristics of the protocol, including (a) communication media, (b) speed, and (c) other protocols such as DNP, assist the seamless communication between devices from different vendors. A security standard called IEC 62351 acts in a supportive manner and implements end-to-end encryption that can prevent the legacy type of assaults, including replay attacks and packet injection.

The work from Yang et al. [140] provided suggestions on potential ways to exploit IEC-104 vulnerabilities to perform attacks against power systems. The authors noticed that the main vulnerabilities stem from the lack of data integrity, availability and authentication mechanisms in the protocol's specification. This opens a window of opportunity for assaults like DoS, replay, spoofing, unauthorized command execution, and buffer overflow.

In [141], Maynard et al. focused on specific IEC-104 vulnerabilities that can lead to MitM attacks. They examined the protocol and also suggested a methodology for performing a replay and a MitM attack. Unlike the rudimentary replay attack, the MitM one requires knowledge of deployment details of the target environment. The vulnerability that renders these attacks possible is the lack of data integrity and authentication measures.

As IEC-104 is popular in the electrical sector, and appears to be a lucrative diode for attackers that wish to inflict damage to this sector. The literature work included in this subsection, highlights on the fact that mainly the lack of authentication and anti-replay mechanisms can make it possible for adversaries to inject malicious or even issue unauthorized commands in the network. The consequences of such an incident can range from a simple local disruption to a cascading effect impacting multiple parts of the power grid.

### D. IEC-61850

IEC 61850 is a modern standard defined under the IEC Technical Committee 57. It consists of a set of communication protocols for devices that are used in electric power systems [142]. The protocol targets interoperability between devices from different vendors along with a common data storage format, also known as Common Data Classes (CDC), that follows a specific hierarchy, namely Physical Devices, Logical Device, Logical Node, Data Object, and Attribute. It also offers security, that is, the IEC 62351-4 and IEC 62351-6 sub-parts than cater for message authentication, channel encryption, and/or cryptographically signed messages to ensure integrity [143]. The protocols used in IEC 61850 are the Manufacturing Message Specification (MMS), Generic Object Oriented Substation Event (GOOSE) and Sampled Measured Values (SMV) [144].

With the help of a custom testbed environment, Yang et al. [145] performed fuzzing kind of attacks towards evaluating the security of the protocols that fall under the IEC-61850 umbrella. The authors observed that the authentication and encryption mechanisms are poorly implemented in certain protocol stacks, which in turn opens a window for MitM, DoS, reconnaissance, injection of commands, modification of data, and spoofing assaults.

Kabir et al. [146] scrutinized the GOOSE protocol of IEC-61850 standard using a properly configured testbed. Specifically, they provided an attack scenario against the protection process of a distribution substation. This can be achieved via the injection of malicious commands and the spoofing of the source address of the devices in the data packets. The authors concluded that the lack of authentication and encryption in these particular implementations, allows the adversary to successfully carry out such an assault.

The work by Silveira and Franco [147] also presented a handful of attacks that originate from vulnerabilities of IEC-61850. They coined these attacks as GOOSE flood, high stNum in GOOSE message, semantic spoofing of GOOSE messages, and GOOSE message replay. Nevertheless, the authors considered implementing only the GOOSE flood attack and elaborated on its mitigation by means of managed switches and software-defined networking (SDN).

IEC-61850 is a relatively new protocol and favorably comes with integrated security features to remedy the issues of its predecessors. Nevertheless, researchers have identified flaws mainly in the way that the protocol is implemented, which in turn may leave room for message spoofing and replay attacks.

### E. PROFINET

PROFINET [148] is an industrial protocol suite standardized and maintained by PROFIBUS & PROFINET International. The current version is an Ethernet-based fieldbus application layer protocol with the real-time capability and is considered a successor to PROFIBUS, which is a classic serial fieldbus. The main advertised perks of PROIFINET are (a) ease of use, (b) integrated diagnostics and safety, (c) scalability for real-time applications, and (d) high availability. The standard supports various topologies, including star, tree, and ring. Note that the protocol also provides a non-real-time mode that is based on UDP/IP. This mode is used for non-time-critical communication, including configuration and diagnostics. No less important, PROFINET is the most used communication protocol for Siemens devices.

Baud and Fester [149] were the first to investigate the potential of mounting MitM attacks on a PROFINET network. They experimented with the Ettercap tool, but the attack was not successful except in the case where the MitM device was placed between the PROFINET controller and the network switch. The authors also mention the possibility of using a "dedicated" device, i.e., a PROFINET device emulator, along with IP and MAC address spoofing to conduct the attack.

Akeberg and Bjorkman [150] elaborated on the feasibility of attacking and gaining control over a PROFINET node using two attacks. In the first, the attacker assaults a PROFINET device with the aim of writing values just before the PROFINET controller does. If successful, the values received next from the controller will be considered duplicates and dropped. The second comprises a MitM to obtain control over the outputs of a device. The attack blossoms out by firstly sending a PROFINET Discovery and basic Configuration Protocol (DCP) identify frame, making all connected nodes to reply with their IP and MAC address. An ARP poisoning phase follows, placing the attacker's device between the controller and the legitimate device. The authors underlined that, especially for the first attack, timing is vital.

Hui and McLaughlin [151] investigated the security issues that possibly exist in newer Siemens PLCs and uncovered vulnerabilities of the PROFINET's discovery protocol, namely DCP. They also detail on a weakness in the protocol's anti-replay mechanism. Precisely, the lack of integrity check in the protocol's acknowledgement packets can lead to either active or passive session hijacking. Moreover, they showed that the device discovery protocol could be violated by forging fake packets, which in turn allows the introduction of rogue PLCs into the network. The operators can be tricked to connect to these rogue PLC using a MitM attack. Since the examined PLCs allow only one ongoing session, the operators will be left without access.

Pfrang and Meier [152] exploited vulnerabilities of PROFINET to conduct two attacks against systems that rely on it. These attacks were based on the vulnerabilities of a) switch port stealing and b) the lack of any authentication measures for the PN-DCP. Specifically, the first vulnerability can cause redirection of commands intended for a particular device or enable the attacker to send malicious commands towards the victim device. The second one can reconfigure the targeted device based on the will of the adversary. Lately, by exploiting a vulnerability in DCP, Mehner and Konig [153] elaborated on a DoS attack on PROFINET. The goal of the attack was to interrupt the Application Relationship (AR) between a PROFINET controller and a device and subsequently obstruct the repair of the system. The previously discovered port stealing attack can interrupt any established AR between the controller and all devices, followed by a reconfiguration attack that exploits the DCP set request, that blocks the operator from restoring the old configuration.

By summarizing the above, it can be discerned that the adversaries can compromise PROFINET protocol communications through simple attacking techniques. Specifically, violation of availability and message authenticity and integrity can be achieved, allowing DoS, MitM and replay kind of attacks. Even the latest version of devices, such as the Siemens S7-1211C PLCs, are prone to this type of attacks as explained in [151].

### F. WirelessHART

WirelessHART is a wireless sensor networking technology based on the Highway Addressable Remote Transducer Protocol (HART), created in 2004 by companies that participate in the HART Communications Foundation (HCF). It operates in unlicensed bands on a mesh topology. Its current version is 2.0 released in 2016 and is defined in IEC/PAS 62591:2016 [154].

The work by Raza et al. [155] elaborated on several vulnerabilities in the WirelessHART protocol. Despite the fact that cryptographic techniques are used to protect the protocol, various attacks are possible. These include packet flooding, gateway spoofing, traffic analysis, resource exhaustion, and desynchronization. Finally, they detail on the use of the Security Manager of WirelessHART that provides the created security keys to the Network Manager, which in turn distributes them to the corresponding devices. The authors pinpointed that the WirelessHART standard lacks of guidelines on how security should be conjuncively applied in the wired part of the network.

Samaddar et al. [156] introduced timing attacks in WirelessHART networks. They elaborated on how such an attack can aid an aggressor in analyzing the eavesdropped traces of the real-time flows of data with the aim of inferring the schedule that those data were exchanged. This is possible due to the fact that the attacker can passively sniff the network traffic without disrupting the regular network flows. If the attackers are capable of monitoring the wireless transmissions in the network, it would require just two hyper periods for them to figure out the communication schedule. Note that the term "hyperperiod" refers to the lowest common multiple of the periods of all the data flows in a network. Finally, the authors proposed a scheme called *SlotSwapper*, which randomizes the schedule over every hyperperiod. This would render the mentioned attack inefficient, and the overall confidentiality of the WirelessHART network will substantially improve.

Since WirelessHART transmits all the information over the air, a lot of attention is given to secure the protocol from eavesdroppers. However, as explained in the context of this subsection, the employed security features are not foolproof. A consequence of that can be the disruption of the exchanged data, causing commotion or interruption to the industrial process.

### G. ZigBee

Zigbee is a wireless protocol maintained by the ZigBee Alliance [157] established in 2002. It enables wireless area network communication in low-power, low data rate, operating in unlicensed bands. Its latest version 3.0, allows increased interoperability along with Internet connectivity in a full protocol stack that utilizes mesh networking and security features. Primarily, Zigbee is considered an IoT protocol. However, due to the latest developments in ICS and especially IIoT, Zigbee is one of the main drivers of this transformation.

In [158] Wright presented a methodology and the corresponding tool for manipulating the distribution of keys in Zigbee protocol with the purpose of decrypting messages or injecting bogus signed and encrypted messages.

Kennedy and Hunt [159], detailed on an association flooding attack that may occur if the coordinator of the Zigbee network does not limit the number of association requests. That is, an attacker can use non-existent devices with a fake address to perform this assault. Furthermore, if a device controlled by the attacker manages to announce itself as a coordinator, it can drop packets and control or disrupt the mesh network formation.

Replay attacks in ZigBee are possible if the participating devices in the network use the same network-wide key as presented by Farha and Chen [160], and other review works [161]. This vulnerability sprang from the incorporation of the frame counter, which was introduced to defend against replay attacks. The latter kind of assaults can also occur when the network key is pre-configured in the devices, and no re-keying is performed. For addressing this issue, the authors propose the use of multiple network keys along with a key generating method.

ZigBee vulnerabilities mainly pertain to the acquisition of the network-wide encryption key by opponents. Typically, for protocols with encryption capabilities, the secret keys comprise a critical asset of the system, and careful consideration must be made regarding their creation, exchange, and storage.

### H. EtherNet/IP

EtherNet/IP is an application layer protocol managed by the Open DeviceNet Vendor Association (ODVA) [162]. It is based on the Common Industrial Protocol (CIP), which represents each device in an industrial network as an object. EtherNet/IP adopts this representation along with Ethernet's standard capabilities. This allows for TCP and UDP connections used to exchange either implicit unicast or multicast messages for time-critical data, or explicit ones for non-time-critical data. EtherNet/IP can be found in great extend in Allen-Bradley devices.

A secure version of EtherNet/IP / CIP has been recently proposed by ODVA with the goal to mitigate the security issues that affect the protocol [163]. CIP Security for EtherNet/IP devices uses the TLS and DLS protocols. Similar to Modbus TCP-Security, this amelioration offers a) authentication of the endpoints via X.509 certificates, b) message integrity and authentication via HMAC, and c) message encryption through AES.

Grandgenett et al. [164] performed an analysis of Allen-Bradley's implementation of EtherNet/IP [165]. The researchers reverse-engineered the protocol to discover any potentially vulnerable packets fields that could be exploited by an aggressor. Their results demonstrate that DoS is feasible through three alternative ways, namely, packet flooding, TCP connection, and session request hoardings. Such attacks can render the controller unresponsive, which in turn will force the operators to reset the device, disrupting the corresponding physical process.

The work of Urbina et al. [166] demonstrated how MitM attacks in EtherNet/IP protocol and related topologies such as the ring topology, can be used to modify sensor measurements and influence actuators in a water treatment testbed. The researchers performed MitM attacks to read data from the EtherNet/IP ring topology, and inject manipulated data in both sensor readings and actuation commands. They also automatically computed the integrity checksum (a non-cryptographic checksum) for the data transmitted via the EtherNet/IP protocol. Lastly, the authors proposed a detection mechanism which relied on sensor measurements vis-à-vis their estimated values, and the ways that the examined attack can deceive this estimation as well.

A fuzzing tool coined *ENIP Fuzz* destined to EtherNet/IP and parts of the CIP was created by Tacliad et al. [167]. Based on three test cases, the creators of the tool managed to identify a new vulnerability that is triggered by an invalid File Type value in the Execute Programmable Controller Communication Commands (PCCC) service of the protocol and can lead to DoS. Actually, PCCC is a service that provides commands which instruct a device to perform actions or respond with status information. The vulnerability was discovered in the protocol implementation of Allen-Bradley's MicroLogix 1100 PLC.

Most of the EtherNet/IP implementations are proprietary, and therefore reverse engineering techniques should be exercised toward discovering flaws in such systems. The vulnerabilities summarized in this subsection can lead to data manipulation and DoS, two of the most severe attacks that can be triggered against an ICS environment.

### I. OPC

Open Platform Communications (OPC) [168], formerly OLE (Object Linking and Embedding) for Process Control, is an open standard protocol used to provide a common communication interface for devices that run specific protocols, including Modbus and DNP3. This allows devices such as a Human-Machine Interface (HMI) that run an OPC client to communicate with various devices that potentially rely on different ICS communication protocols via an OPC server. The server acts as a "middle" man to apply the appropriate hardware communication protocol translations. Implementations of the older version of OPC referred to as OPC Classic, were only supported by the MS Windows OS family. The most recent version, OPC Unified Architecture (OPC-UA), is also available outside the Microsoft ecosystem.

The work of Qi et al. [169] extends the work of Wang et al. [170] by focusing on detecting flaws on OPC, Distributed Component Object Model (DCOM), and Remote Procedure Call (RPC) parts of the ecosystem. Their tool, called *OPC-MFuzzer*, considered two distinct classes of vulnerabilities, namely insecure API calls and non-sufficient validation of inputs. For the RPC related test cases, a simple Python script replicated the functionality of a DCOM and created the desired RPC requests. The execution of the tool provided information about already known vulnerabilities, including remote code execution and DoS, but also revealed some novel ones that were reported privately to the corresponding vendors.

Puys et al. [171] scrutinized the security of the OPC-UA protocol through formal analysis methods. Specifically, they employed the *ProVerif* tool to perform cryptographic protocol verification. They identified flaws in the way the free OPC implementation *FreeOpcUa* [172] performs the cryptographic signing of messages, as well as authentication issues that can lead to unintended access to the system.

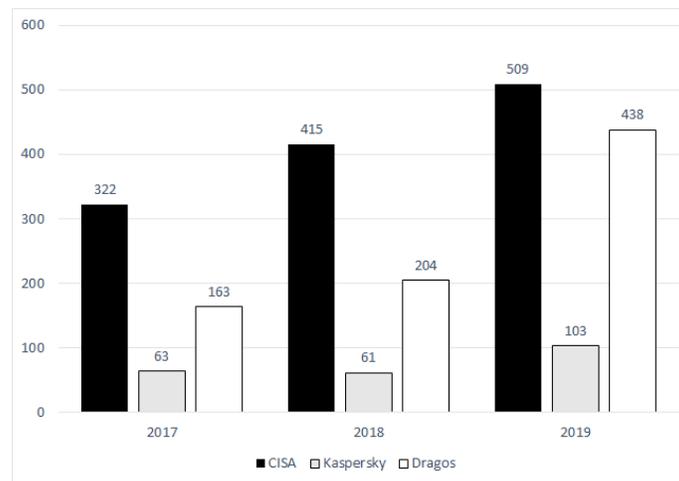

Figure 12: Number of vulnerabilities identified by each organization

The work by Roepert et al. [173], demonstrates various methods that can be used to discover vulnerabilities in OPC-UA servers. The researchers specifically attempted to penetrate into such servers by first discovering them in the network, and then by testing the OPC-UA authentication for weak credentials. Once they got access, they obtained all the required information to check for potential vulnerabilities. They discovered that some of the tested OPC-UA servers are prone to DoS attacks.

We observe that OPC is prone to common attacks that are met in legacy IT environments. Insecure API calls and insufficient validation along with specific product implementation errors can provide the adversaries with a handful of knowledge since the nature of OPC is to interconnect diverse ICS components. Therefore, not only the availability of services can be compromised, but also the confidentiality of key exchanged data, including "secret recipe" ones transferred among devices.

### J. Discussion

With reference to the previous subsections, it becomes obvious that the vulnerabilities in ICS protocols are introduced by either the antiquated assumption that the systems of interest are isolated or from implementation errors or oversights in the protocol stack. That is, given that availability is typically the primary protection goal for automation systems, most automation protocols do not consider security as a design tenet, but mainly as an afterthought. Even more, as nowadays, most of these protocols have shifted from the serial world and the siloed, isolated model to the Ethernet realm, they are directly susceptible to legacy Internet attacks. And of course, by design, Ethernet does not offer any kind of protection for any automation protocol.

On the other hand, implementation errors, such as those exposed by the work of Yang et al. [145], demonstrate that the pressure which is applied to vendors due to the highly competitive market, impels many asset owners to hastily adopt potentially vulnerable protocols in their systems. Both these issues are difficult to tackle given that (a) vendors and relevant stakeholders, including CISA need to first discover and then patch any exposed vulnerability, (b) opposite to legacy IT systems, the handling of ICS equipment is not a trivial process, because often even a small-scale update requires the halt of the physical operation, and (c) in some sectors, including electrical grid and nuclear plants, there are legal implications regarding the patching of firmware, as the affected systems must be re-certified by the responsible authorities [174], [175].

In fact, the number of vulnerabilities reported by ICS-CERT, Kaspersky and Dragos [176], [177], [178], [179], [180], [181], increase every year as demonstrated in Figure 12. Naturally, this inflation is attributed to more thorough testing, and not due to potential downgrade in the quality of the systems.

In the past, where the systems were isolated from each other, there was no need for confidentiality and integrity mechanisms to be implemented for the ICS protocols. Nowadays, standardization committees and other related parties strive to ameliorate the security of the relevant protocols in every possible aspect and at the earliest opportunity or next scheduled revision. Characteristically, the attempts from Modbus and DNP to introduce a wrapper (Modbus over TLS) and an integrated protection solution (DNP3-SA), correspondingly, are sitting in this mindset. However, sometimes, the ambiguity of protections offered by a protocol can significantly impact the use of specific features of it and lead to a false sense of security. A prominent example of this situation is the selective use of authentication for specific function codes in DNP3-SA, that, as indicated in [125], resulted in the lowering of the overall security level of the system. Another one, is the use of a network-wide key in ZigBee protocol implementation that is pre-configured in some devices, and no re-keying is performed [160].

The latter is an indicator that even in the move to IIoT, security problems still arise and therefore, security experts and asset owners should understand the new kinds of threats and attack surfaces, and take the appropriate actions in mitigating the potential effects of attacks. Cloud integration to ICS can intensify the problem if this integration is not properly configured. For this reason, asset owners should securely transmit their data to their on-premises servers in addition to cloud storage. The non-mission-critical data can be accessed via the cloud infrastructure and used in a supplemental way, entirely outside the ICS.

Another approach for the asset owners is to apply their own protections based on legacy security technologies for the sake of reducing the attack surface of OT systems. Some of the advantages and disadvantages of these two angles, namely legacy protection schemes vs. integrated solutions, are presented in [182]. More analytically, traditional approaches, including VPN/IPSec and TLS, have the advantage of being well-tested and thoroughly examined by many researchers over the years. On the downside, they typically introduce significant latency to the system and are associated with key management issues as a rule. On the other hand, integrated solutions such as those in DNP3-SA and IEC-61850, have minimal impact on the performance of the system and are cost-efficient. However, these solutions have not received enough and thorough testing and examination. Therefore, they cannot compare with industry-standard cryptographic methods in terms of maturity.

Besides the protection of the ICS specific protocols, security measures must also be applied to legacy networking equipment, including switches, routers and their matching protocols. When feasible, MAC/SYN flooding protections, ARP spoofing protections, encrypted traffic and access control lists, must be present. Furthermore, the strategic placement of firewalls and IDS can assist in mitigating threats that in the presence of an exploitable vulnerability can be weaponized and penetrate multiple levels of an ICS architecture. Finally, where applicable and in conjunction with digital defenses, physical security such as locking cabinets, Close-Circuit Television (CCTV) and port lock-ins, must also be properly enforced.

## VII. ICS DEVICE VULNERABILITIES

This section is dedicated to describing vulnerabilities specific to ICS devices. These devices operate primarily at the lower levels of the Purdue model, as illustrated in Figure 1. In this section, we intentionally focus on academic works that discover weaknesses and prove the feasibility of possible attacks. Often, vulnerabilities against ICS equipment and sometimes their accompanied software are discovered and disclosed by companies and individuals outside academia. Some of these organizations provide significant resources for the identification of such vulnerabilities. CISA [183] creates and maintains a list of identified vulnerabilities, including the severity level based on the CVSS base score; the CI sectors then proceed to potential mitigation actions. While alternative methods of exploitation have been uncovered, we can organize corresponding works as ones that: (a) perform reverse engineering, (b) aim in the unauthorized modification of control logic, (c) ladder-logic based, (d) seek the installation and execution of malware, natively, at the ICS device level, (e) ones that stem from lack of proper authentication mechanisms of the executed control logic, or finally the ones that (f) achieve leakage of sensitive data or disruption of device operation through the abuse of side-channels.

### A. Reverse Engineering

Typically, the main challenge of generating PLC malware from an attacker's viewpoint is that full knowledge of the PLC operational aspects is required. This includes a holistic understanding of the interfaces with the connected devices, which can typically be obtained only via espionage. An additional domain-specific challenge revolves around the closed-source nature of the compilation process. While reverse engineering can have many benign applications, it is always considered as the first step of malicious actions, and its automation may significantly reduce the cost of malware payload construction.

The methodology introduced in [184] and subsequently the tool in [185] aim to automate the malicious payload construction process for PLCs. More often than not, even after rigorous analysis, the main difficulty from an attacker's viewpoint falls down to determining the operational details of PLCs, such as the exact memory locations that are responsible for regulating the controlled devices. The presented tool coined SABOT receives a high-level specification of device's behavior as input. That document is compiled by the adversary and contains definitions regarding temporal logic properties of devices. The tool then retrieves the benign control logic bytecode from the target PLC and automatically identifies mappings of PLC memory locations to physical devices. To successfully complete the decompilation phase, the control logic bytecode is converted to an intermediate set of constraints C variables. Then, these are translated into the process model using the modeling language of the NuSMV model checker. In this way, the tool is able to automatically modify a generic malicious payload defined by the attacker into one that is capable of infecting the target PLC. This work proved that assailants could manipulate the operations of specific devices, in this case, Level 0 devices, i.e., sensors and actuators, with just a superficial understanding of the physical process.

Similarly, Keliris and Maniatakos [186] present a framework for automatically reverse engineering full PLC binaries with the aim of reconstructing the complete Control Flow Graph (CFG) of the control logic. The proposed framework consists of the following phases: (a) reverse engineering of the corresponding binary format to identify details such as the header section and subroutine delimiters, (b) create a knowledge base that contains information such as the mappings of the IO to memory addresses, and (c) perform the analysis. Still, the bottleneck of the proposed framework lies in the first phase, where a significant amount of time must be dedicated by a human expert to understand the (possibly undocumented) binary formats and the specific to the platform compiler conventions.

A decompiler for the ladder logic, Laddis is presented in [187]. Laddis can decompile a program on the fly by observing packets that contain control code transmitted during the PLC configuration cycles. It then interprets the binary ladder logic and decompiles it into a human-readable ASCII form. In a subsequent step, the ASCII output gets transformed into the graphical ladder logic. The decompilation process starts by identifying the rung structures in the binary file. The binary format is complex and not well documented, but once fully dissected and understood it is easy to create automated tools that perform the translation process. For example, the authors identified that each rung always starts with two bytes of zeros, followed by two bytes containing a signature of the rung. Once again, the process requires intense manual labor, and it is platform-specific, but it is reusable when completed once.

As part of the control logic attack implementation presented in [188] the authors contributed a decompiler referred to as Eupheus. The decompiler is designed to produce Instruction List (IL) source code from binaries specific to the RX630 platform. The decompiler consists of a database component that contains a large number of mappings for various types of instructions, such as including input and output, relative branch, and function block, among others. The database was populated by a human expert after manual labor. The mapper component identifies RX630 instructions in the code block and performs a lookup in the database component to obtain the corresponding IL instructions.

JEB [189] is a commercial decompiler that supports the Siemens PLCs platforms. Typically, the Siemens PLC programs are written in either Statement List, Structured Control Language, or Ladder Logic. Irrespective of the source language, the PLC program is compiled to MC7 machine code and packed into a binary blob that constitutes block data elements, metadata description, and machine code. The decompiler can parse all these blocks, and the MC7 machine code is disassembled and decompiled to a language similar to C code.

To date, reverse engineering of ICS binaries remains a costly procedure that can only be achieved by well-funded and highly skilled attackers. To the best of our knowledge, existing implementations are either platform-specific or partial, and no generic or fully automated tool for the task is widely available.

*B. Control Logic Injection & Modification Attacks*

The control logic installed in a PLC specifies how that device will control aspects of the physical process. It is defined by the EW by engineers and gets "downloaded" to the PLC. However, the control logic can be subject to malicious alterations, primarily because PLCs do not support sophisticated cryptographic means for verifying its authenticity and authorship, e.g., via digital signing, or simply because such mechanisms, despite being available, are usually neglected.

Three alternative attacks are presented in [187] that alter the control logic of PLCs to degrade the integrity and availability of the system. In the first methodology, after injecting malicious control logic to a target PLC, the attacker acts as a MitM to remove the infected code from the packets that get produced when the engineering software attempts to verify the running PLC logic. In this way, the opponent is able to effectively conceal any traces of malicious alteration of the control logic by presenting the original (uninfected) logic to the investigator. In the second assault, the attacker acts as a MitM to replace a selected number of control logic instructions with arbitrary instructions (e.g., "noise" as sequences of 0xFF bytes), which in several cases ends up crashing the programming software. In the last methodology presented, the attacker installs malicious control logic crafted in a way that forces the engineering software to paralyze every time it attempts to obtain the control logic from the PLC. In this scenario, the attacker can perform the malicious substitution of ladder logic once and leave the ICS network.

Kalle et al. [188] were driven by the observation that most academic works assume that exploitation of the engineering software as a first step to infect the control logic of the PLC. This, of course, takes for granted that a pre-existing vulnerability in that software exists and that the attacker has detailed knowledge of the type and version of the software in the targeted environment. The proposed CLIK attack consists of four main steps: (a) the attacker first compromises the PLC directly (e.g., by bypassing rudimentary authentication mechanisms such as short passwords) and acquires the control logic (in binary format) from the PLC, (b) the attacker decompiles the retrieved binary on the fly and then alters it by injecting the malicious instructions, (c) as a next step, the attacker "downloads" the altered version of the control logic back to the PLC, (d) finally, the methodology assumes the operation of a virtual-PLC, a "fake" device that continuously reports the original, benign version of the control logic back to the engineering software in an effort to conceal the fact that any alteration took place. Nevertheless, the presented attack still assumes that the attacker has successfully penetrated into the OT network and is able to alter messages on the network. While there are alternative ways to achieve this, the latter is still a strong assumption.

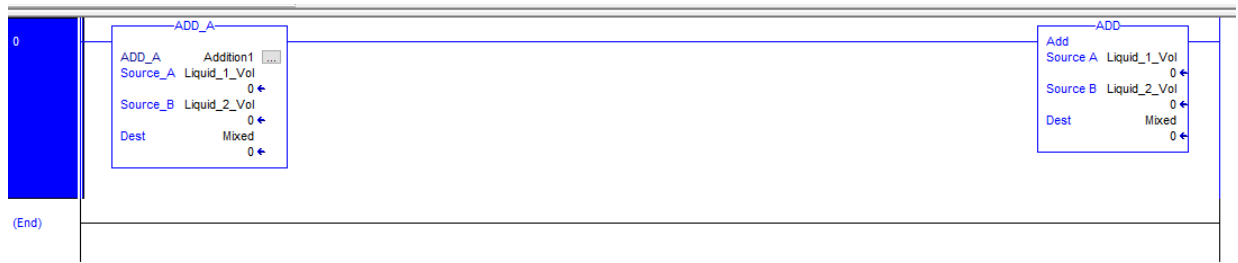

Figure 13: Two instructions that look similar in first place.

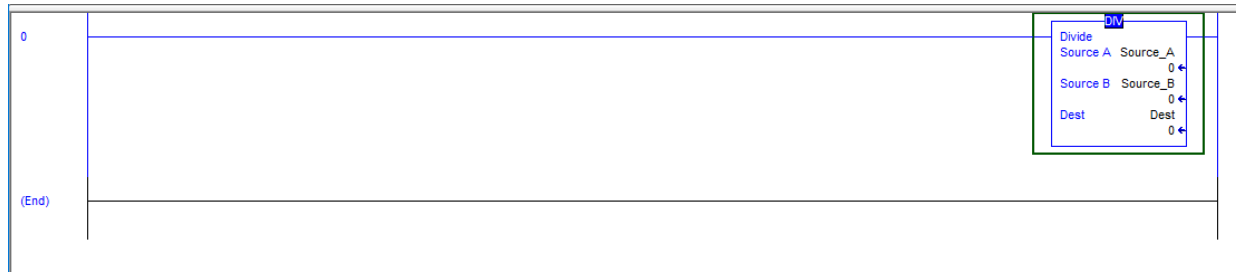

Figure 14: The malicious logic of the concealed "Add-on" instruction that performs a DIV instead of an ADD

Yoo and Ahmed [190] explored various methods of injecting malicious logic to a PLC directly through the network without requiring physical access to a PLC. They describe two alternative attack methodologies, namely, (a) Data Execution, (b) Fragmentation with Noise Padding. In the first type, malicious code is transferred to the data blocks of a PLC. By default, these sections are primarily used to hold the sensor measurement values. The examined PLCs do not support data execution prevention (DEP) mechanisms, thus allowing the execution of arbitrary code directly from these locations. The Fragmentation and Noise Padding attacks aim in evading detection of deep packet inspection IDS. Such techniques tend to perform poorly when analyzing small packets (smaller than a pre-defined n-gram size). Each "write" request fragment sent to the PLC is crafted to contain only one byte of malicious code concatenated with random "noise" sequence of bytes. However, the attacker crafts the address field of the next fragment so that it overwrites the noise data of the previous write request when stored in the PLC memory. When all the fragments have been received, the malicious payload will be fully reconstructed at the PLC side.

### C. Ladder Logic Based Attacks

Ladder logic is one of the IEC61131-3 compatible languages for programming control logic in PLCs. It is a visual programming language that represents the execution flow of a PLC as a circuit diagram of vertical rails and a series of horizontal rungs between them. It mostly targets factory engineers and technicians that are not fluent in traditional programming languages. It is by far the most popular way to define control logic, with popular alternatives being the C-like format of *structured text* or the assembly-like format of *instruction list*.

Govil et al. [191] introduced the concept of ladder logic bombs (LLBs), i.e., malicious snippets of ladder logic that may be implanted in the benign logic by a malicious engineer with direct access to the EW. LLBs execute the malicious sequence of instructions to disrupt the normal operations only after being triggered by a specific event. The authors classify the LLBs into different categories according to their effect, namely, (a) LLBs that may lead to DoS, (b) LLBs to manipulate sensor readings and commands, and (c) LLBs for stealthy logging of data. The novelty of the proposed approach lies in the way modifications are crafted. Specifically, LLBs aim to evade the attention of a human analyst who wishes to verify the integrity of the logic. In a typical example, a wary engineer would obtain the current control logic from PLC devices and manually inspect it for possible alterations. For small-scale logic and for experienced engineers who have a good understanding of the physical processes, it may be feasible to analyze the code and identify malicious blocks. Naturally, manual analysis is proven inefficient for complex, large-scale scenarios. An example of such practice is to perform a different action by hiding the malicious logic inside a user-defined "Add-on" instruction. The reader should bear in mind that ladder logic is a visual programming language, and user-defined instructions can be visually very similar to the traditional but more frequently used "ADD" block. A similar example can be seen in Figures 13 and 14.

The work of Serhane et al. [192] focuses on ladder logic code vulnerabilities or simply bad code practices that may become the cause of bugs or be exploited by attackers. Indeed, today most ladder logic coding standards tend to be company/sector-oriented and aim towards the optimal operation of the physical processes. Bad coding practices may include (a) reusing certain unsafe operands multiple times in a single program e.g., a timer, (b) the use of instructions that contain user-defined logic but can easily remain unnoticed in the overall ladder logic program, for example, a FIFO instruction may perform operations such as logging critical information, (c) using empty branches that may lead to bypassing critical execution blocks, (d) misplaced operands within the same code/rung that can generate race conditions with unpredictable consequences, (e) no detailed alarms and poorly implemented code diagnostics which may allow runtime errors to pass unnoticed, and (f) including unused tags or operands can be leveraged by attackers that can trigger unauthorized actions. The authors also underline that keeping the PLC in program mode opens the door to adversaries to inject malicious logic into the PLC.

### D. Native ICS Malware

The majority of the PLC infecting malware capitalizes on vulnerabilities of the EW or other platforms that are based on commodity hardware and software, i.e., PCs running MS Windows OS. Several works, however, proved that it is possible to create malware that operates directly at the PLC side.

Spenneberg et al. [193] demonstrated the first Proof of Concept (PoC) worm written in structured text that propagates among PLCs without the involvement of an engineering workstation. This can be achieved due to inadequate security measures such as the lack of integrity protection in the PLC and the default settings of the access protection (by default, it is turned off). Unlike most approaches, the malware does not modify the benign control logic, but rather executes the malicious code in parallel. First, it scans the network for additional target PLC devices, and then it replicates itself and transmits its code to discovered targets. The researchers managed to identify the structure of each message by analyzing the anti-replay mechanism and the POU that transfers the program to the PLC. In this way, they were able to circumvent the anti-replay mechanism to store previously sent messages in a static internal database (DB) POU. In subsequent phases, these messages were altered to include the worm and replayed to spread the malware. The worm itself, has the capability to perform DoS by injecting an infinite loop in each PLC. Furthermore, it can survive the restart of the device. The main assumption is that the worm is first introduced to the OT network through another PLC infected in a previous stage.

The work from Garcia et al. [194] presents HARVEY, a rootkit that gets installed to the device's firmware and has the capability to inspect the control logic and then modify its instructions. The rootkit is also aware of the control process that the PLC handles and can intercept the measurement inputs that are used by this process. The rootkit is stealthy in its nature as the operators cannot detect its presence using the conventional PLC monitoring capabilities. The real novelty of the malware is that the malware has knowledge of the physical process that the PLC controls. Indeed, changing input and output values may raise suspicion and lead to detection through the effects. For example, if the malware would cause an unjustified increase of pressure in a pipe, a sensor could detect such inconsistency and automatically set an independent safety mechanism. Instead, Harvey implements a model of the target system and ensures that the malicious modifications remain within thresholds of expectancy.

Generally, native PLC malware is implemented at the firmware level. Yet, firmware level modification should not be considered trivial. In many cases, the firmware can be updated only through direct physical access, say, via an SD card. Furthermore, unlike control logic software, the firmware is typically protected from unauthorized updates via cryptographic means such as digital signatures.

### E. Unauthorized Access

Beresford [195] identified several vulnerabilities regarding the Siemens Simatic S7 PLCs. These vulnerabilities can be used by malicious actors to perform: (a) replay attacks, (b) authentication bypass, (c) DoS, (d) remote memory dumps, and (e) remote shell access. The reader should take into account that the sessions between a client and the PLC never expire, so they can be recorded and replayed to bypass any basic authentication mechanism, at any time. For example, the lack of anti-replay mechanisms made it possible for an adversary to capture commands such as CPU START and STOP and replay them back to the PLC. The DoS can be performed using forged packets that are not handled correctly and thus result in rebooting PLC even when a control process is live, something that in turn might lead to dangerous operational conditions. Memory dumps can be exploited by an attacker to acquire information about industrial processes. Such can be altered and sent back to the PLC. Last but not least, since all services execute using root privileges, malicious payloads can be created and inserted into the PLC to gain full access to the system.

Klick et al. [196] demonstrate how an attacker could extend access to all of PLC in the production network and, under the circumstances, the corporate IT network. They make the assumption that if an internet-facing PLC has no authentication enabled, the behavior of an EW can be mimicked and malicious control logic can be injected to the PLC. The first stage of their attack is to inject an SNMP Scanner written in the Statement List (STL) programming language that runs in addition to the normal control code. After acquiring the results of the SNMP scan, the adversaries can inject a SOCKS Proxy along with the control logic, thus enabling them to reach more devices in the local network. The developed scanner and proxy can be downloaded on a PLC without service interruption to the already-existing PLC program, hindering the infection from the operators. To automate the steps of the attack, they also provide a tool called PLCinject.

The work from Wardak et al. [197] investigates the issues in access control mechanisms of PLCs. The research focuses on the available access controls that exist on the S7-400 PLC, one of the high-end devices Siemens offers. The researchers note that there are three levels of access control, (a) no protection, (b) write protection and (c) read/write protection. The last two require password authentication, while the "no protection" level does not require any authentication at all. They captured packets during the password authentication exchange to determine the utilized encoding/encryption scheme. Using automated analysis, they determined that a rudimentary, custom XOR-based transformation is applied to protect the passwords. Thus, the researchers could easily infer the passwords and gain access to the PLC.

Keliris et al. [198] discuss a vulnerability discovered in the authentication mechanism of several protective relays of the General Electric (GE) Multilin protection and control family of products. The identified vulnerability allows an attacker to gain remote or local access and obtain weakly encrypted user passwords, which in turn allows unauthorized access to the device. The vulnerability stems from the fact that passwords are encrypted using a weak, custom encryption algorithm. By unleashing a Chosen Plaintext Attack (CPA) against the employed encryption algorithm, the researchers managed to acquire the corresponding plaintext password and subsequently access the protective relay.

An investigation of authentication protocols used by PLCs, is presented by Ayub et al. [199]. In this work, the security posture of devices from Schneider Electric, Allen-Bradley Automation Direct, and Siemens is examined by leveraging the produced network traffic. The researchers unveiled that the design issues and vulnerabilities in authentication mechanisms are (a) lack of nonce, (b) small-sized encryption key, (c) weak encryption scheme, (d) client-side authentication, (e) reuse of the same keys in multiple sessions, (f) improper session management, (g) sensitive information disclosure, and (h) no write access protection. Finally, all the vulnerabilities are mapped to the MITRE ATT&CK framework to verify whether these are exploitable.

The security mechanisms that are designed to prevent unauthorized access to a large portion of ICS devices can be easily circumvented. This is a great risk to the ICS network even if alternative security controls are in place. As potential countermeasures, more secure cryptography methods such as digital signatures can be applied to alleviate the current situation.

*F. Side Channel Analysis*

Side-channel attacks rely on capitalizing analog emissions such as electromagnetic, thermal, or acoustic signals that get involuntarily transmitted during the regular operational cycles of devices. Such signals are usually ignored or treated as noise, but in reality, they are highly correlated with certain operations executed by the subject device. For this reason, they may provide valuable information to a passive observer, which progressively may lead to leakage of sensitive information such as cryptographic keys. Other types of side-channel attacks may include timing and cache attacks. The former leverages timing differences in accessing data or completing operations upon data that lead to making assumptions about the data itself. The latter type takes advantage of the fact that most CPU architectures allow resource sharing of certain levels of cache among the software that is being executed by the CPU. These practices have been applied successfully in high-end systems such as servers, but devices involved in the ICS realm appear to be an even more alluring target since (a) their CPUs are slower and follow a less complex architecture, and (b) the supported software is simple and the corresponding operations are performed in an infinite loop.

Krishnamurthy et al. [200] described the possibility for malware to rely on acoustic emissions of actuators, e.g., that of a motor controlling a valve as part of a closed-loop process, towards creating a covert channel. This secret transmission had zero impact on the stability, performance, and signal characteristics of the feedback control loop. The attack assumes that an acoustic sensor (microphone) is introduced in close proximity to the device. The exfiltrated information can be used to launch further attacks that can either determine spoofed sensor values that can possibly degrade the system performance or even shut down the process. Remarkably, it is demonstrated that a eight character password can be leaked in around two minutes, a 32-bit control setpoint parameter in around one minute, and a 128-bit key in a little over four minutes.

Tychalas and Maniatakos [201], examine the applicability of cache timing side-channel attacks against PLC devices. Modern PLC designs have software components at various operational levels which make eviction-based and speculation-based (e.g., Spectre) strategies feasible for unleashing side-channel attacks. For example, an Evict-and-Reload attack is presented as a candidate for retrieving process data from the I/O module; the same technique may lead to leakage of cryptographic material from the OS, while a Spectre-like attack may result in leaking data from network or processes running at the application level. The study was conducted considering systems that use the Codesys framework [202] but is primarily theoretical and no experimental evaluation was provided.

Blinkware [203] is an attack that achieves information leakage among embedded systems through the use of the optical side-channel. In the described example, sensitive information is transmitted via memory-mapped peripherals such as LED by copying data from arbitrary memory locations via the DMA controller. DMA enables peripherals to access the memory directly without any interference of the CPU. More specifically, in this attack, the LEDs achieve transmission of information by manipulating the blink rate and brightness of the LEDs. One of the advantages of relying on DMA is that since it imposes a minimal burden to the CPU, traditional ways of anomaly detection by profiling the CPU performance may be rendered ineffective. Similarly the Waterleakabe malware [204] relies on the optical side-channel to achieve transmission of the sensor readings, from lamps that are connected to the digital output of the PLC to a compromised video recorder camera placed one meter away.

### G. Discussion

This section includes academic works that describe PoC attacks and vulnerabilities which could potentially be abused in future incidents. The vulnerabilities analyzed may lead to theft of information, and loss of view (LOV), control (LOC), and safety. As mentioned previously, these devices reside in the ICS environments for a prolonged time, and even if security holes are discovered, the asset owners face update challenges or the cumbersome incorporation of other security measures that should mitigate the vulnerabilities.

Some incorrect assumptions regarding the improvement of the security posture that exist in the industry are (a) the ICS are not connected to the Internet; therefore, attacks are limited to the physical realm, (b) that security can be achieved by not disclosing information about the implementation details (security through obscurity), and (c) the sense (illusion) of absolute security that common defence techniques provide, including the use of firewalls, antivirus software, and encryption.

Even if ICS devices are not connected to the Internet, they provide some networking capabilities used for monitoring and maintenance. In this respect, the adversary needs to only acquire access to the appropriate network to unleash attacks, and possibly exfiltrate information or cause control process disruption or destruction.

The use of custom encoding instead of cryptographically secure processes as identified in [193] and in [199] falls under one of the above categories. With moderate efforts the attackers can circumvent those measures and gain access to the equipment. Even worse, there are cases where the credentials are hardcoded or insufficiently protected, so their retrieval becomes a trivial process [205], [206]. Moreover, the update process of ICS devices should include adequate and well defined policies that verify the new system in an effort to avoid attacks such as those given in [207] and [194].

Another threat origins from smart devices that can be incorporated into ICS, forming an IIoT environment. For example, attackers can use vulnerable sensors as an entry point into ICS and damage critical components. Remotely exploitable vulnerabilities such as the Ripple20 [208], and the Urgent/11 [209] in the Trecks' and VxWorks' TCP/IP stacks respectively, can have severe repercussions to the control processes. This gives an example that asset owners and vendors should be aware that potential vulnerabilities in IIoT devices can be leveraged both as a backdoor and final payload against ICS.

Besides the mitigation of vulnerabilities that stem from erroneous design and implementation in ICS devices (similarly to those identified in [192]), contributions such as the "Top 20 Secure PLC Coding Practices Project" [210], presents an effort to use PLC-specific features to achieve security. This can be accomplished by developing PLC programs in such way that can offer knowledge about input validation, physical integrity checks, and introduction of field devices.

Finally, as mentioned in subsection VI-J, physical security that can be achieved with the use of locking cabinets, CCTV and port lock-ins, can also be applied to protect devices that are vulnerable to attacks that can be only performed with physical presence in the environment.

### VIII. CONCLUSION

The once isolated and monolithic CI systems, including electricity, water, gas, manufacturing, and transportation, have evolved into an increasingly complex and interlinked System of Systems, potentially exposing a very fertile attack surface. Even more, paradigms such as IIoT and the 5G communication technology are already culminating in a tighter union of systems. In fact, certain processes in smart-buildings, smart-cities, transport systems, including autonomous vehicles, and factories are already remotely controlled via the internet. This means that security analysts should not isolate themselves in silos by solely concentrating on threats and vulnerabilities of specific sectors. Instead, a cross-sectoral cyber thinking and defense approach is needed.

This work explored the so far most important ICS and CI security incidents and scrutinized their key aspects. It also elaborated on the common factors and vulnerabilities that enabled real life incidents and detailed potential mitigation measures. A broad discussion on the relevant network protocols and key infrastructure components, typically met in such realms, has also been provided. From this analysis, it becomes apparent that the use of antiquated network protocols and design inefficiencies in ICS devices beget vulnerabilities, which in turn leave room for disruptive and even persistent attacks. Given the long system lifecycles, the required engineering resources, and vulnerabilities for which patches are unlikely to be applied (because they are considered legitimate features of the technology or product architecture by the vendors), contemporary security mechanisms cannot always be straightforwardly administered.

Without any doubt, the ICS and CI are vital to a well-functioning modern society. Therefore, instead of focusing on specific malware or tactics used by the adversaries, organizations should create countermeasures and mitigation steps to deter and prevent the early stages of any potential attack (or at least the stages prior the physical process is affected). Defenders must identify the malicious behavior and the interactions that can occur inside their environment, thus creating efficient and vigorous defense mechanisms for each control process. In addition, particular focus should be given to lessen the impact of successful attacks by concentrating the efforts on the aspects of human behavior, but also on the reliability and resiliency of the control processes. No less important, the use of non-intrusive security solutions for ICS is another research area that is worthy of consideration. Such systems offer security in a way that it does not alter or affect the system's performance to the point of adversely impacting any part of interest.